\documentclass[3p,sort&compress]{elsarticle}

\usepackage{graphicx}
\usepackage{amsfonts}
\usepackage{amsmath}
\usepackage{amssymb}
\usepackage{dsfont}
\usepackage{indentfirst}
\usepackage{color}
\usepackage[caption=false]{subfig}
\usepackage{bm}

\newcommand{\beq}{\begin{equation}}
\newcommand{\eeq}{\end{equation}}
\newcommand{\beqa}{\begin{eqnarray}}
\newcommand{\eeqa}{\end{eqnarray}}
\newcommand{\bpmat}{\begin{pmatrix}}
\newcommand{\epmat}{\end{pmatrix}}
\newcommand{\bmat}{\left(\begin{array}}
\newcommand{\emat}{\end{array}\right)}

\newcommand{\myfrac}[2]{\leavevmode\kern.1em\raise.5ex\hbox{\scriptsize
$#1$}\kern-.1em {\scriptsize /}\kern-0.10em\lower.25ex\hbox{\scriptsize $#2$}}

\renewcommand{\Re}[1]{\hbox{\rm Re}\left(#1\right)}         
\renewcommand{\Im}[1]{\hbox{\rm Im}\left(#1\right)}         
\newcommand{\e}{\,\hbox{\rm e}}                                      
\newcommand{\vect}[1]{{\bm #1}}                                       

\begin{document}

\title{Low-loss surface modes and lossy hybrid modes in metamaterial waveguides}

\author{Benjamin R. Lavoie\corref{cor1}}
\ead{brlavoie@ucalgary.ca}

\author{Patrick M. Leung}
\ead{pmyleung@ucalgary.ca}

\author{Barry C. Sanders}
\ead{sandersb@ucalgary.ca}

\cortext[cor1]{Corresponding author}

\address{Institute for Quantum Information Science, University of Calgary, Alberta, T2N 1N4, Canada}
\begin{abstract}
We show that waveguides with a dielectric core and a lossy metamaterial cladding (metamaterial-dielectric guides) can support hybrid ordinary-surface modes previously only known for metal-dielectric waveguides. These hybrid modes are potentially useful for frequency filtering applications as sharp changes in field attenuation occur at tailorable frequencies. Our results also show that the surface modes of a metamaterial-dielectric waveguide with comparable electric and magnetic losses can be less lossy than the surface modes of an analogous metal-dielectric waveguide with electric losses alone. Through a characterization of both slab and cylindrical metamaterial-dielectric guides, we find that the surface modes of the cylindrical guides show promise as candidates for all-optical control of low-intensity pulses.
 \end{abstract}


\maketitle


\section{Introduction}

Electromagnetic metamaterials are materials designed to produce an electromagnetic response that is not possible with just the constituent materials alone~\cite{Balmain:2005}. Within the past decade metamaterials have been of interest~\cite{Boardman:2005,Ramakrishna:2005}, with metamaterials that operate at optical frequencies under current development~\cite{Shalaev:2007,Boltasseva:2008,Xiao:2009}. Potential uses include cloaking devices~\cite{Cai:2007}, and perfect lenses~\cite{Pendry:2000}. Given that the electromagnetic response of a metamaterial is, in principle, tailorable, there has been considerable interest in developing metamaterial waveguides~\cite{Shadrivov:2003,Halterman:2003,Qing:2005, He:2008,D'Aguanno:2005}. The modes supported by metamaterial waveguides exhibit a wide range of properties with applications to improving the design of optical and near ultra-violet systems. We investigate the properties of modes in both slab and cylindrical waveguides with a lossy metamaterial cladding and show how both waveguide geometries are able to support low-loss surface modes.

By exploiting effects that arise from surface plasmon-polaritons propagating along a single interface~\cite{Kamli:2008,Moiseev:2010}, a metamaterial-dielectric waveguide is able to support low-loss surface modes. The low-loss modes in a metamaterial-dielectric waveguide display a lower attenuation than is possible for the same mode in a metal-dielectric waveguide, thereby opening up the possibility of low-loss plasmonic waveguides. The low-loss surface modes we observe arise due to the same interference effects that cause the near-zero-loss surface mode observed on a single metamaterial-dielectric interface~\cite{Kamli:2008}. The single-interface waveguide is used to confine the fields in order to enhance optical nonlinearity for all-optical control at a single-photon level~\cite{Moiseev:2010}. Rather than using a single-interface waveguide for all-optical control, employing a low-loss surface mode of a cylindrical waveguide, which confines the fields in the transverse direction, could enhance the optical nonlinearity that is required for all-optical control of pulses.

Energy loss, which is a major obstacle in the implementation of metamaterial waveguides, is present in metamaterials due to scattering and heating effects in the materials used to construct them~\cite{Tassin:2012}.  The artificial structure of metamaterials alters the scattering and heating effects of the materials~\cite{Penciu:2010}. To fully understand the properties of the modes in a metamaterial waveguide, it is crucial to include loss in the model. So far little consideration has been paid to the effects of loss~\cite{D'Aguanno:2005,He:2005}. In order to model energy loss, we allow both the permeability and permittivity of the metamaterial to be complex-valued and frequency-dependent. For permeability we use a Lorentz-like model that describes the behavior of metamaterials designed for optical frequencies~\cite{Penciu:2010}, whereas the standard Drude model is used for permittivity of the metamaterial. 

We compute dispersion, attenuation, and effective guide width for various modes of a cylindrical guide and, for comparison purposes, a slab guide. We show that a metamaterial-dielectric waveguide supports both surface and ordinary TM modes as well as hybrid ordinary-surface TM modes. Hybrid ordinary-surface modes are supported in metal-dielectric parallel-plate waveguides~\cite{Barlow:1973}. Hybrid modes are not predicted by lossless models. We find that a metamaterial-dielectric guide is able to support hybrid modes over a much wider frequency region than is seen in metal-dielectric guides. Some modes have attenuation curves with sharp changes over narrow frequency regions. The frequency at which this change occurs depends on the structural parameters, thereby enabling a metamaterial-dielectric guide to act as a frequency filter.


\section{Theory}

To characterize light in a waveguide, we solve the Helmholtz equation in each region (core and cladding) for both the electric field, $\vect{E}(x,y,z;t)$, and the magnetic field, $\vect{H}(x,y,z;t)$:
\beq
\nabla^2 \vect{F}(x,y,z;\omega)=-\omega^{2}\epsilon(\omega)\mu(\omega)\vect{F}(x,y,z;\omega),\label{generalwaveeq}
\eeq
where $\vect{F}(x,y,z;\omega)$ is either $\vect{E}(x,y,z;\omega)$ or $\vect{H}(x,y,z;\omega)$. We are interested in traveling plane-wave solutions, which have the form
\beq
\vect{E}(x,y,z;\omega)=\vect{E}(x,y)\e^{i(k_z z-\omega t)}
\eeq
and
\beq
\vect{H}(x,y,z;\omega)=\vect{H}(x,y)\e^{i(k_z z-\omega t)}.
\eeq
for $k_{z}$ the complex propagation constant. As the permittivity and permeability of the claddings are complex functions, so too is the propagation constant $k_z:=\beta+i\alpha$ with $\alpha$ the attenuation per unit length for a propagating field and $\beta$ the propagation wavenumber. Both $\beta$ and $\alpha$ are real and positive. A negative value for $\alpha$ implies the material is active (exhibits gain) whereas negative $\beta$ corresponds to fields propagating in the negative $z$ direction. Once the solutions to Eq.~(\ref{generalwaveeq}) in each region are found, we apply the appropriate boundary conditions at each interface. Applying the boundary conditions yields the dispersion relation for the waveguide, which is geometry-specific.

As we are interested in lossy waveguides, permittivity and permeability of the metamaterial must both be complex-valued and are denoted $\epsilon=\epsilon'+i\epsilon''$ and $\mu=\mu'+i\mu''$, respectively, for $\epsilon'$, $\epsilon''$, $\mu'$, and $\mu''$ all real, and $\epsilon''>0$ and $\mu''>0$. The expressions for $\epsilon$ and $\mu$ are determined by how metamaterials are constructed~\cite{Penciu:2010}. A common approach used in metamaterial design for optical frequencies is to exploit the naturally occurring electric response of metals to achieve a negative permittivity and design the structure of the metamaterial to generate a magnetic response that gives a negative permeability~\cite{Boltasseva:2008}. 

The natural electric response of metals to an electromagnetic field is described by a frequency-dependent expression for the permittivity given by the Drude model~\cite{Sernelius:2001}:
\beq
\frac{\epsilon(\omega)}{\epsilon_0}=1-\frac{\omega_{\textrm{e}}^2}{\omega\left(\omega+i\Gamma_{\textrm{e}}\right)},\label{epsilonMM}
\eeq
where $\omega_{\textrm{e}}$ is the plasma frequency of the metal and $\Gamma_{\textrm{e}}$ is a damping constant. The structurally induced magnetic response of a metamaterial takes the form~\cite{Penciu:2010}
\beq
\frac{\mu(\omega)}{\mu_0}=1+\frac{F \omega^2}{\omega_0^2-\omega\left(\omega+i\Gamma_{\textrm{m}}\right)},\label{muMM}
\eeq
where $F$, a function of the geometry, is the magnetic oscillation strength, $\omega_0$ is a binding frequency, and $\Gamma_{\textrm{m}}$ is the damping constant for the magnetic interactions. The magnetic response is generated by resonant currents that are driven by the electromagnetic field. For parameters that are real and positive, the real part of both Eqs.~(\ref{epsilonMM}) and (\ref{muMM}) can be either positive or negative whereas the imaginary parts are strictly positive.

The index of refraction of any lossy material is in general complex and given by
\beq
n=\pm\sqrt{\frac{\epsilon\mu}{\epsilon_0\mu_0}}=n_{\rm{r}}+i n_{\rm{i}},\label{complexn}
\eeq
where the sign is chosen such that $n_{\rm{i}}\geq 0$ to ensure the material is passive (gain-free). The other case, $n_{\rm{i}}< 0$, corresponds to a material that exhibits gain; i.e., the intensity of the fields increases as the fields propagate through the material. The choice of the sign in Eq.~(\ref{complexn}) to ensure $n_{\rm{i}}>0$, means that at frequencies where both $\epsilon'<0$ and $\mu'<0$, the metamaterial behaves as a negative-index material with $n_{\rm{r}}<0$. 

The frequency region where $\epsilon'$ and $\mu'$ have opposite signs (one negative and one positive) correspond to metamaterials that do not allow propagating waves. Many metals at optical frequencies act as electric plasmas and are characterized by $\epsilon'<0$ and $\mu'>0$; hence, for metamaterials that are similarly characterized, we use the term metal-like to describe them. The metal-like region is what D'Aguanno et al.\ call the opaque region~\cite{D'Aguanno:2005}. For most frequencies in the metal-like region $n_{\rm{i}}>\left|n_{\rm{r}}\right|$. In contrast, metamaterials with $\epsilon'>0$ and $\mu'<0$ are not metal-like per se but would correspond to materials that display electromagnetic responses consistent with an effective magnetic plasma~\cite{Ramakrishna:2005}. For frequency regions with both $\epsilon'>0$ and $\mu'>0$, the metamaterial simply behaves as a dielectric with a frequency-dependent refractive index.

\begin{figure}[t,b,p] 
        \centering
	\includegraphics[width=0.5\columnwidth]{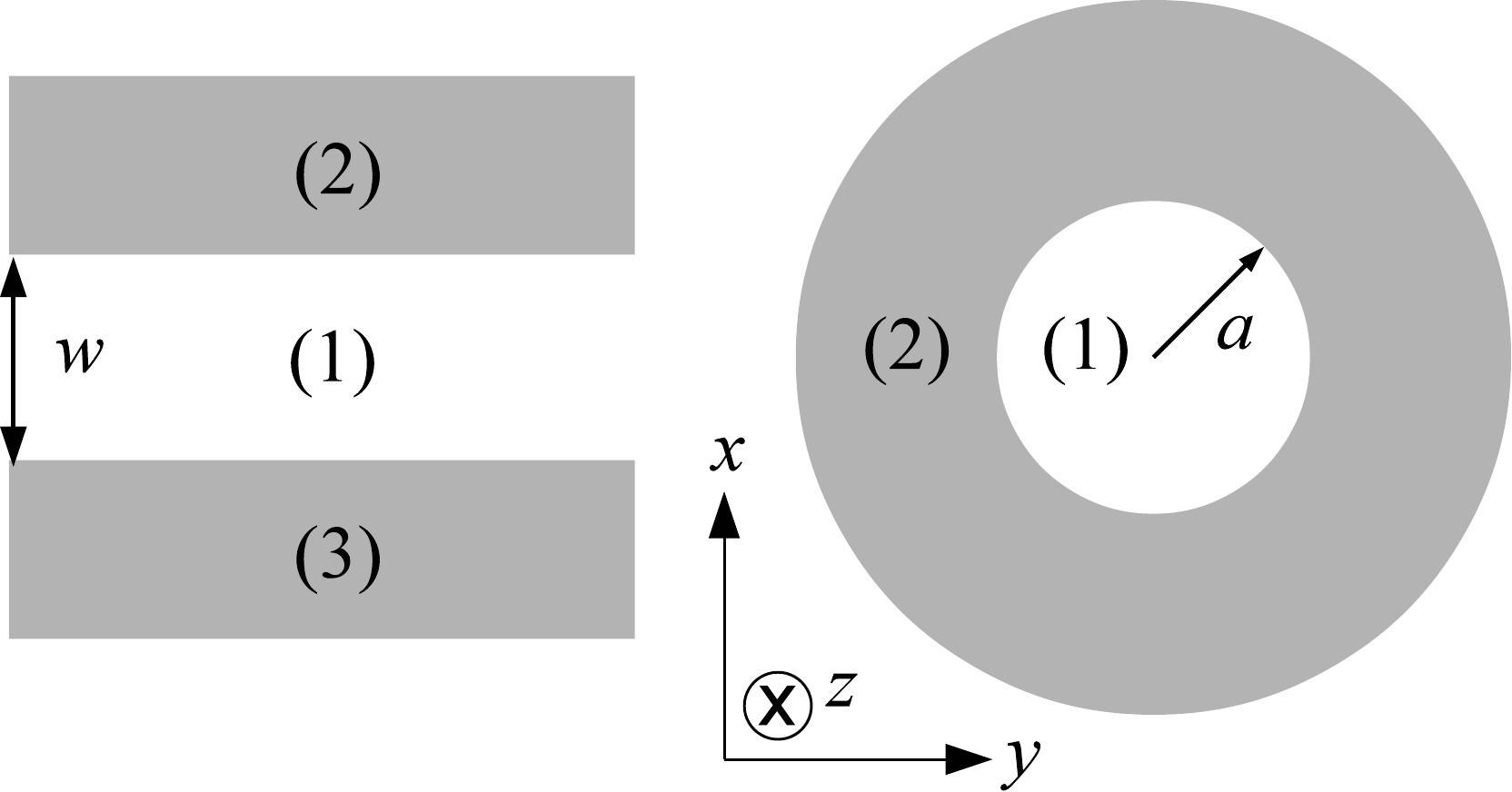}
	\caption{A diagram of the structure of the slab and cylindrical waveguides. Region (1) in both guides is the core and is a lossless dielectric, whereas regions (2), and region (3) of the slab guide are the cladding layers and are metamaterial. The width of the slab guide is $w$ and the radius of the cylindrical guide is $a=w/2$. Fields propagate in the $z$ direction (into the page), whereas the $x$ and $y$ directions are the transverse directions.}
     \label{fig:guidediagram}   
\end{figure}

We begin by studying the slab geometry as it has features in common with both a single interface (no transverse confinement) and cylindrical waveguides (supports multiple types of modes). Additionally, the slab waveguide has been studied for various materials~\cite{Yeh:2008,Kaminow:1974}, which allows us to check our methods against accepted results. The slab waveguide consists of three layers as seen in Fig.~\ref{fig:guidediagram}.
The center layer is the core, and the width of the waveguide, $w$, is the thickness of the core. We choose the width of the slab guide to be $w=4\pi c/\omega_{\rm{e}}$, such that the frequency at which $w$ is equal to the wavelength of the field is near the middle of the frequency range in which the metamaterial is metal-like. The core is surrounded on both sides by cladding. The cladding layers are sufficiently thick to substantially contain the evanescent fields, as the effective width of the guide (the width of the core plus the skin depth at the interface) is typically only slightly larger than $w$. The permittivity and permeability of the cladding material is described by Eqs.~(\ref{epsilonMM}) and (\ref{muMM}) whereas they are constant in the core.

The slab guide supports two distinct classes of modes, transverse magnetic (TM) and transverse electric (TE). The TM modes have non-zero magnetic field components only in the transverse direction ($H_{z,x}=0$) whereas the TE modes have electric field components which are non-zero only in the transverse direction ($E_{z,x}=0$). The existence of TM surface modes in our system is partially dependent on the cladding having $\epsilon'<0$ and $\mu'>0$, whereas the TE modes require $\epsilon'>0$ and $\mu'<0$ in the cladding. As the expressions for $\epsilon$ and $\mu$, Eqs.~(\ref{epsilonMM}) and (\ref{muMM}) respectively, are different, we cannot expect TE surface modes to simply be a generalization of TM surface modes.  

We choose to study the TM modes as the forms of the response functions allow for the existence of TM surface modes for a large set of parameters, whereas the TE surface modes are not as robust. Applying the boundary conditions to the solutions of Eq.~(\ref{generalwaveeq}) for the TM modes leads to the dispersion relation
\beq
\frac{\epsilon_1}{\gamma_1}\left(\frac{\epsilon_2}{\gamma_2}+\frac{\epsilon_3}{\gamma_3}\right)=-\left(\frac{\epsilon_1^2}{\gamma_1^2}+\frac{\epsilon_2 \epsilon_3}{\gamma_2\gamma_3}\right)\tanh\gamma_1 w, \label{slabdisp1}
\eeq
which must be satisfied for the existence of a propagating mode~\cite{Yeh:2008}. Here
\begin{equation}
\gamma_j:=\sqrt{k_z^2-\omega^2\epsilon_j\mu_j}\label{transversewavenumber}
\end{equation}
are the complex wave numbers for the transverse components of the fields for $j=1,\,2,\,3$ refering to the core and two cladding layers respectively. As Eq.~(\ref{slabdisp1}) is transcendental, we solve numerically for the complex propagation constant $k_z$.  To obtain the dispersion relation for the TE modes, simply replace any explicit appearance of $\epsilon_j$ in Eq.~(\ref{slabdisp1}) by $\mu_j$.

Cylindrical guides enable transverse confinement of the modes but are of interest to us for more than this reason. The cylindrical geometry is well studied in other contexts and is used in a number of applications, e.g.,\ fiber optics. Additionally, as for the slab geometry, the dispersion relation for a guide with cylindrical geometry can be obtained analytically. The cylindrical waveguide shown in Fig.~\ref{fig:guidediagram} consists of a dielectric cylindrical core of circular cross-section with radius $a$. A cladding layer, which has a frequency dependence described by Eqs.~(\ref{epsilonMM}) and (\ref{muMM}), surrounds the core. As with the slab geometry, the cladding layer is assumed to be arbitrarily thick.

We again assume traveling-wave solutions, but we must include all components of both electric and magnetic fields. In general the modes of a cylindrical guide are not TE or TM but in-between modes that contain all of the field components and are called HE or EH~\cite{Yeh:2008}. The names HE and EH allude to the fact that all components of both electric and magnetic fields are non-zero. The TM and TE modes of the cylindrical guide are special cases of the general solution. Using the definitions
 \beq
J'_m\left(a\kappa_1\right)=\left.\frac{\partial J_m\left(r \kappa_1\right)}{\partial r}\right|_{r=a}\nonumber
\eeq
and
\beq
K'_m\left(a\gamma_2\right)=\left.\frac{\partial K_m\left(r \gamma_2\right)}{\partial r}\right|_{r=a}{\textstyle ,}\nonumber\\
\eeq
where $ J_m$ and $K_{m}$ are the Bessel function and modified Bessel function, respectively, the dispersion relation for the cylindrical guide is~\cite{Yeh:2008}

\beq
\hspace{-1.25cm}\frac{k_z^2m^2}{a^2\omega^2}\left(\frac{1}{\kappa_{1}^{2}}+\frac{1}{\gamma_{2}^{2}}\right)^2=\left(\frac{\mu_1}{\kappa_{1}^{2}}\frac{J'_m\left(a \kappa_1\right)}{J_{m}\left(a\kappa_1\right)}+\frac{\mu_2}{\gamma_{2}^{2}}\frac{K'_m\left(a\gamma_2\right)}{K_m\left(a\gamma_2\right)}\right)\left(\frac{\epsilon_1}{\kappa_{1}^{2}}\frac{J'_m\left(a \kappa_1\right)}{J_{m}\left(a\kappa_1\right)}+\frac{\epsilon_2}{\gamma_{2}^{2}}\frac{K'_m\left(a\gamma_2\right)}{K_m\left(a\gamma_2\right)}\right)\label{cyldisprel}.
\eeq

Here $\gamma_{2}$ is the complex wave number in the radial direction for the cladding and is given by Eq.~(\ref{transversewavenumber}), and $\kappa_1=i\gamma_{1}$ is that for the core, whereas  $m$ is an integer characterizing the azimuthal symmetry. With the dispersion relations for both geometries, we are able to determine the propagation constants for the allowed modes.


\section{Modes}

The metamaterial-dielectric slab waveguide supports a number of TM modes with three distinct mode behaviors, namely ordinary, surface and hybrid modes. Each mode may display more than one behavior, with different behaviors supported for different frequency regions. The mode behavior is determined by the relative sizes of the real and imaginary parts of $\gamma_{1}$, the wavenumber perpendicular to the interface inside the core. To simplify the discussion we use ordinary wave, surface wave, and hybrid wave when referring to the character, or behavior, of a mode and use the term mode or explicitly name the mode (e.g.\ TM$_{j}$) when we are discussing a specific solution to the dispersion relation. 

We label the modes in a manner consistent with lossless waveguides~\cite{Yeh:2008}:\ a numeric subscript $j$ refers to ordinary waves and indicates how many \emph{zeros} (how many times the $H_{y}$ field changes sign) the mode has inside the core, whereas subscripts ``s'' and ``a'' indicate the symmetric and antisymmetric surface modes, respectively. Symmetric and anti-symmetric refers to the symmetry present in the $H_{y}$ field. We extend the labelling scheme to the modes of lossy waveguides even though the defining features may not be present for all frequencies.

Ordinary waves are caused by traveling electromagnetic fields inside the core reflecting from the core-cladding interface(s). The reflected fields overlap with the incident fields causing an interference effect, with constructive interference corresponding to the allowed modes. The condition for ordinary-wave behavior is $\Im{\gamma_{1}}>> \Re{\gamma_{1}}$, which ensures that, in the slab guide, the fields have an oscillating wave pattern in the transverse direction of the core.

Surface waves are the result of electrons at the core-cladding interface carrying energy along the waveguide. The energy from the fields is transferred to the electrons causing them to oscillate, providing a medium for energy transport. This type of collective oscillation of electrons at a surface coupled with electromagnetic waves is called a surface polariton or surface plasmon-polariton~\cite{Nkoma:1974}. The polariton field of the surface mode decays exponentially as a function of the distance from the interface. Mathematically the condition for surface waves is $\Re{\gamma_{1}}>> \Im{\gamma_{1}}$, which causes the intensity to be concentrated around the interface(s) between the core and cladding and a relatively small intensity through most of the core. In slab guides only two surface wave TM modes are supported, namely symmetric (TM$_{\rm{s}}$) and anti-symmetric (TM$_{\rm{a}}$). 

In frequency regions where both $\Re{\gamma_{1}}$ and $\Im{\gamma_{1}}$ are comparable, the mode is a hybrid wave. Hybrid waves can be understood as a product of two features in the transverse direction, namely an evanescent feature and an oscillatory feature. To see how the two features combine, consider the components of the magnetic field in the core of the the slab waveguide, which have the form
\beqa
H&=&H_{0}\e^{\gamma_{1}x}\e^{i(k_{z}z-\omega t)}\nonumber\\
&=&H_{0}\e^{\gamma'_{1}x}\e^{i\gamma''_{1}x}\e^{i(k_{z}z-\omega t)},\label{hybridfield}
\eeqa
where $H_{0}$ is a constant, the complex wavenumber $\gamma_{1}=\gamma'_{1}+i\gamma''_{1}$, with $\gamma'_{1}$ and $\gamma''_{1}$ both real, and the terms $\e^{\gamma'_{1}x}$ and $\e^{i\gamma''_{1}x}$ represent surface-wave and ordinary-wave behavior, respectively. The hybrid waves are then a combination of surface waves, i.e., fields being carried by electron oscillations, and ordinary waves due to fields traveling in the core and reflecting off the interfaces thereby resulting in interference effects.

For hybrid waves a substantial portion of the energy is transferred along the interfaces as well as in the core, so hybrid waves are like a combination of ordinary waves and surface waves. Hybrid waves are not seen in dielectric-dielectric guides and only for a very narrow frequency range of the TM$_{\rm{a}}$ mode in metal-dielectric guides, where the mode changes from a surface wave to an ordinary wave.

The metamaterial-dielectric cylindrical guide supports the same three mode types as the slab guide, though they differ slightly as the cylindrical guide has a circular symmetry. The fields of the ordinary wave TM modes in the cylindrical guide have the oscillating wave pattern of a vibrating circular membrane.  The fields in the cladding decay exponentially, as $\Re{\gamma_{2}}>> \Im{\gamma_{2}}$. The modes of the cylindrical guide are labeled similarly to the slab guide, with the transverse magnetic modes labeled as TM$_{j}$. For the cylindrical guide however, a numeric value of $j$ refers to the number of \emph{zeros} in $H_{y}$ that lie between the center of the core and the core/cladding boundary. 

The HE modes have two indices (HE$_{m,j}$), where the azimuthal symmetry is characterized by the integer $m$, which determines the number of oscillations in the fields over an angle of $2\pi$. The index $m$ here is the same as that in Eq.~(\ref{cyldisprel}).  The index $j$ plays the same role for HE modes as it does for TM modes. Due to the circular symmetry of the cylindrical guide it supports only one surface wave TM mode. However, the cylindrical guide supports a number of surface wave HE modes of different orders with the symmetry determined, as with the ordinary wave TM modes, by the azimuthal parameter $m$. Modes of the cylindrical guide with surface-wave behavior are denoted by the index ``s'' for both TM and HE (e.g. TM$_{\rm s}$, HE$_{1,\rm s}$.


\section{Characterization}

With the physical nature of the three wave types supported by a metamaterial-dielectric guide understood, we now characterize the waveguides. To characterize the two guide geometries we examine how the effective refractive index, attenuation along the propagation direction, and effective guide width for a variety of supported modes vary with frequency. The following values are used for the metamaterial parameters~\cite{Kamli:2008}: $\omega_{\rm e}=1.37\times10^{16}{\rm s}^{-1}$, $\Gamma_{\rm{m}}=\Gamma_{\rm e}=2.73\times10^{13}{\rm s}^{-1}$, $\omega_0=0.2\omega_{\rm e}$, $F=0.5$ whereas the core is described by $\epsilon_1=1.3\epsilon_{0}$ and $\mu_1=\mu_{0}$.

\begin{figure}[t,b] 
      \centering
	\subfloat{\label{fig:epsmu}\includegraphics[width=0.45\columnwidth]{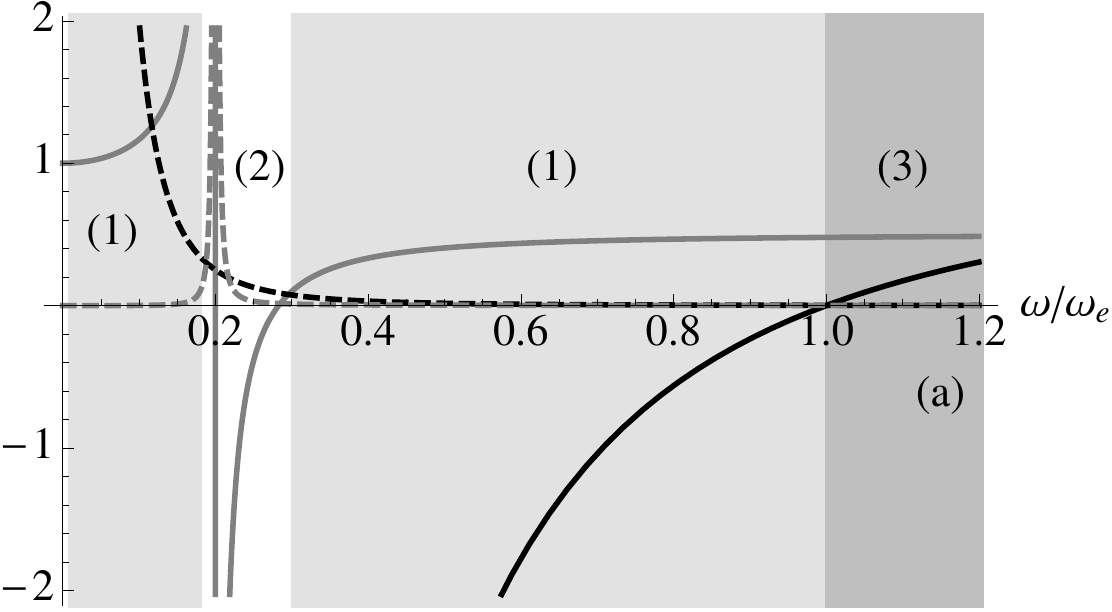}}
	\subfloat{\label{fig:nimmindex}\includegraphics[width=0.45\columnwidth]{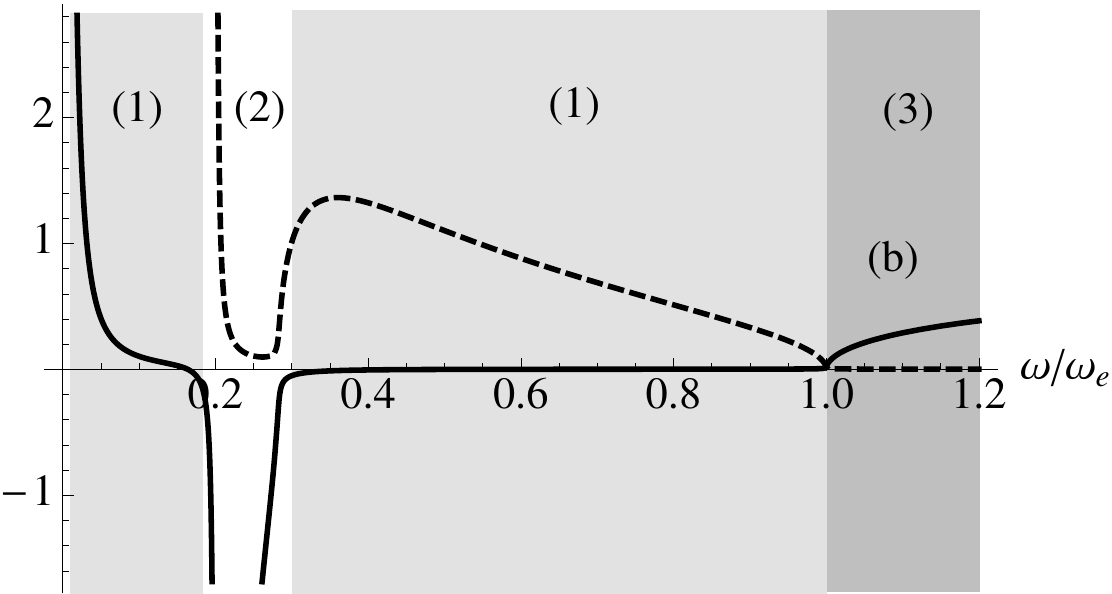}}
	\caption{Plots of the real parts (solid lines) and imaginary parts (dashed lines) of \protect\subref{fig:epsmu} $\epsilon/\epsilon_{0}$ (black) and $\mu/\mu_{0}$ (gray), and \protect\subref{fig:nimmindex} the refractive index of the metamaterial as a function of frequency. Regions (1) in each graph are the metal-like regions, region (2) is the negative index region and region (3) is the dielectric region.}
     \label{fig:nimmregions}   
\end{figure}

Frequency regions with differing refractive index characteristics for the metamaterial are identifiable (Fig.~\ref{fig:nimmregions})
when the above parameters are used along with Eqs.~(\ref{epsilonMM}) and (\ref{muMM}). Above $\omega_{\rm{e}}$ the metamaterial behaves as a lossy dielectric with both $\epsilon'>0$ and $\mu'>0$. There are two \emph{metal-like} regions, one for $0.3\omega_{\rm{e}}\lesssim\omega<\omega_{\rm{e}}$ and one for $\omega\lesssim0.2\omega_{\rm{e}}$. We say \emph{metal-like} because the regions are characterized by $\epsilon'<0$ and $\mu'>0$, which is typical for metals at optical frequencies but not seen in dielectrics. In the frequency range $0.2\omega_{\rm{e}}\lesssim\omega\lesssim 0.3\omega_{\rm{e}}$ the metamaterial has a negative-index. The binding frequency, $\omega_{0}$, in Eq.~(\ref{muMM}) causes rapid changes in the refractive index (Fig.~\ref{fig:nimmregions}\subref{fig:nimmindex}) around $\omega=0.2\omega_{\rm e}$. 

We characterize the slab geometry first, and, to provide a basis for comparison, we also include results for a metal-dielectric waveguide. We assume the slab guide is symmetric, meaning all of the parameters are identical for both cladding layers. The parameters we use to describe the metal are the same as for the metamaterial, with the exception that $\mu_{2,\,3}=\mu_{0}$. To characterize the behavior of the modes we compare the real and imaginary parts of $\gamma_{1}$.
\begin{figure}[t,b] 
     \centering
      \subfloat{\label{fig:relgammaSlabOrd}\includegraphics[width=0.45\columnwidth]{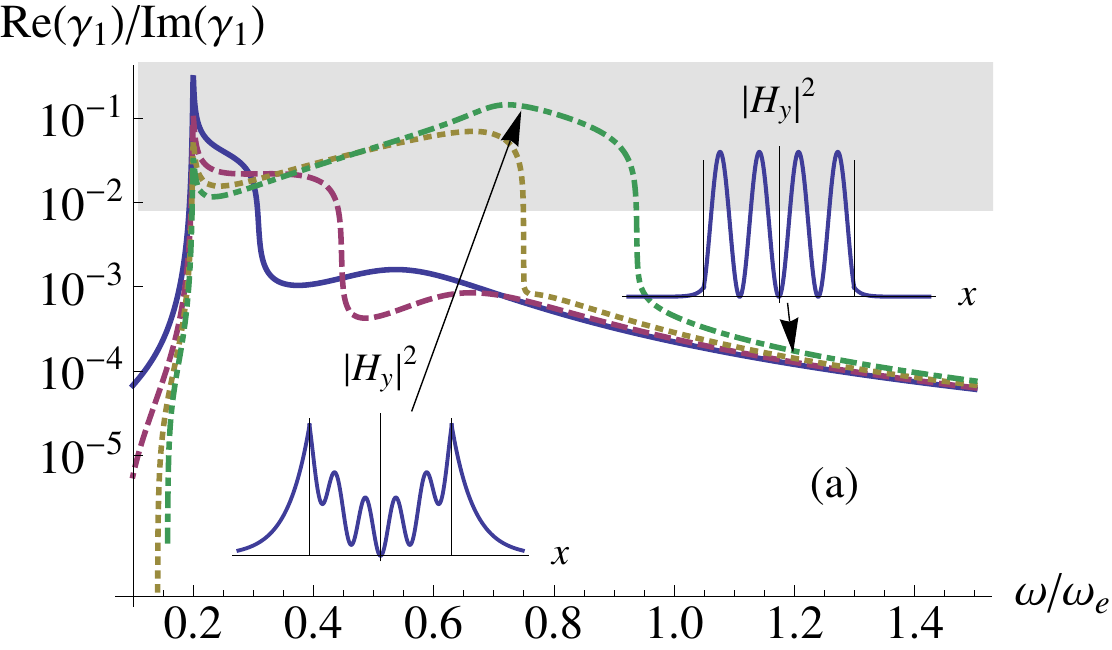}}
      \subfloat{\label{fig:relgammaSlabSurf}\includegraphics[width=0.45\columnwidth]{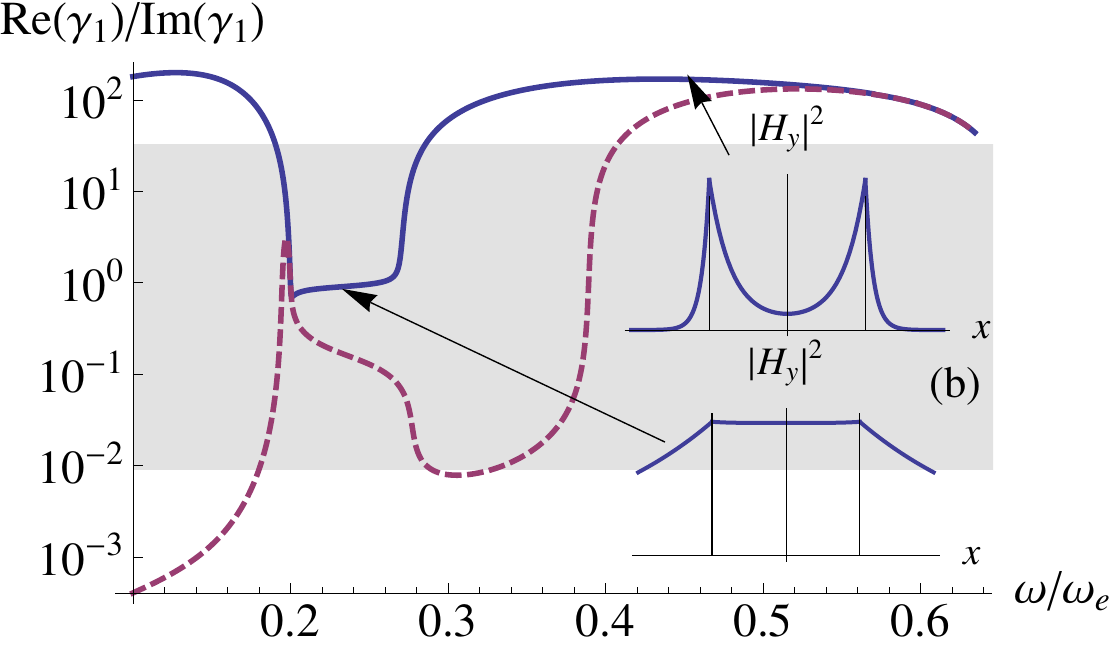}}
     \caption{Plots of $\Re{\gamma_{1}}/\Im{\gamma_{1}}$ for \protect\subref {fig:relgammaSlabOrd} the TM$_{0}$ (solid), TM$_{1}$ (dashed), TM$_{2}$ (dotted), and TM$_{3}$, (dot-dashed) and \protect\subref{fig:relgammaSlabSurf} the TM$_{\rm{s}}$ (solid) and TM$_{\rm{a}}$ (dashed) modes of the metamaterial-dielectric slab guide. The shaded region indicates the nominal range of hybrid waves, as there is no well-defined boundary. The insets are plots of $|H_{y}|^{2}$, which show the transverse profile of the mode at select frequencies. The thin vertical bars in the insets indicate the core/cladding boundary.}\label{fig:relgammaSlab}   
\end{figure}
A plot of $\Re{\gamma_{1}}/\Im{\gamma_{1}}$ (see Fig.~\ref{fig:relgammaSlab}) shows where the real and imaginary parts of $\gamma_{1}$ are of comparable magnitude, indicated by shaded regions, which correspond to hybrid waves. Outside the shaded regions, either the real or imaginary part is dominant and the mode is a surface or ordinary wave, respectively. The insets of the plots also show the transverse profile of select modes at frequencies that lie in regions of different mode behavior. 

The transverse mode profile does not always present a clear distinction between ordinary, surface, and hybrid modes. As noted earlier, the fields of ordinary waves are generally distributed evenly across the core and have an oscillatory pattern, whereas surface modes are characterized by the fields being concentrated at the core/cladding boundary and decaying toward the center of the core. Hybrid waves, however, have characteristics of both wave types. One distinguishing feature of hybrid waves is a large amount of the field is present in the cladding layer (fat tails). The two insets for hybrid waves in Fig.~\ref{fig:relgammaSlab} clearly show the fat tails indicative of hybrid waves.

Four modes are plotted in Fig.~\ref{fig:relgammaSlab}\subref{fig:relgammaSlabOrd} showing the regions of ordinary and hybrid-wave behavior for each. Hybrid-wave behavior is present for $\Re{\gamma_{1}}/\Im{\gamma_{1}}\gtrsim10^{-2}$, which translates to the TM$_{0}$ mode for $0.2\omega_{\rm e}\lesssim\omega\lesssim0.3\omega_{\rm{e}}$, the TM$_{1}$ mode for $0.2\omega_{\rm e}\lesssim\omega\lesssim0.45\omega_{\rm{e}}$, the TM$_{2}$ mode for $0.2\omega_{\rm e}\lesssim\omega\lesssim0.75\omega_{\rm{e}}$ and the TM$_{3}$ mode for $0.2\omega_{\rm e}\lesssim\omega\lesssim0.95\omega_{\rm{e}}$. The insets are plots of $|H_{y}|^{2}$ for the TM$_{3}$ mode and show the difference in the transverse profile between the ordinary and hybrid waves. The inset for $\omega\approx1.2\omega_{\rm{e}}$ is an ordinary wave ($\Re{\gamma_{1}}/\Im{\gamma_{1}}\approx10^{-3}$), whereas the inset for $\omega\approx0.8\omega_{\rm{e}}$ is a hybrid wave ($\Re{\gamma_{1}}/\Im{\gamma_{1}}\approx10^{-1}$).

Figure~\ref{fig:relgammaSlab}\subref{fig:relgammaSlabSurf} shows the TM$_{\rm{s}}$ and TM$_{\rm{a}}$ modes of the metamaterial-dielectric slab guide. Hybrid-wave behavior is present for $10^{-2}\lesssim\Re{\gamma_{1}}/\Im{\gamma_{1}}\lesssim50$, which translates to the TM$_{\rm s}$ mode for  $0.2\omega_{\rm e}\lesssim\omega\lesssim0.3\omega_{\rm{e}}$ and the TM$_{\rm{a}}$ mode for $0.18\omega_{\rm e}\lesssim\omega\lesssim0.4\omega_{\rm{e}}$. From the plots, we see that the TM$_{\rm s}$ and TM$_{\rm a}$ modes are surface waves for frequencies corresponding to $\Re{\gamma_{1}}/\Im{\gamma_{1}}\gtrsim50$. The insets in this figure are plots of $|H_{y}|^{2}$ for the TM$_{\rm{s}}$ mode and show that when $\Re{\gamma_{1}}/\Im{\gamma_{1}}\gtrsim50$ the mode is a surface wave (fields concentrated at core/cladding boundary), whereas it is a hybrid wave when $\Re{\gamma_{1}}/\Im{\gamma_{1}}\lesssim50$.

\begin{figure}[t,b] 
       \centering
       \subfloat{\label{fig:TMSlabDispBOUND}\includegraphics[width=0.45\columnwidth]{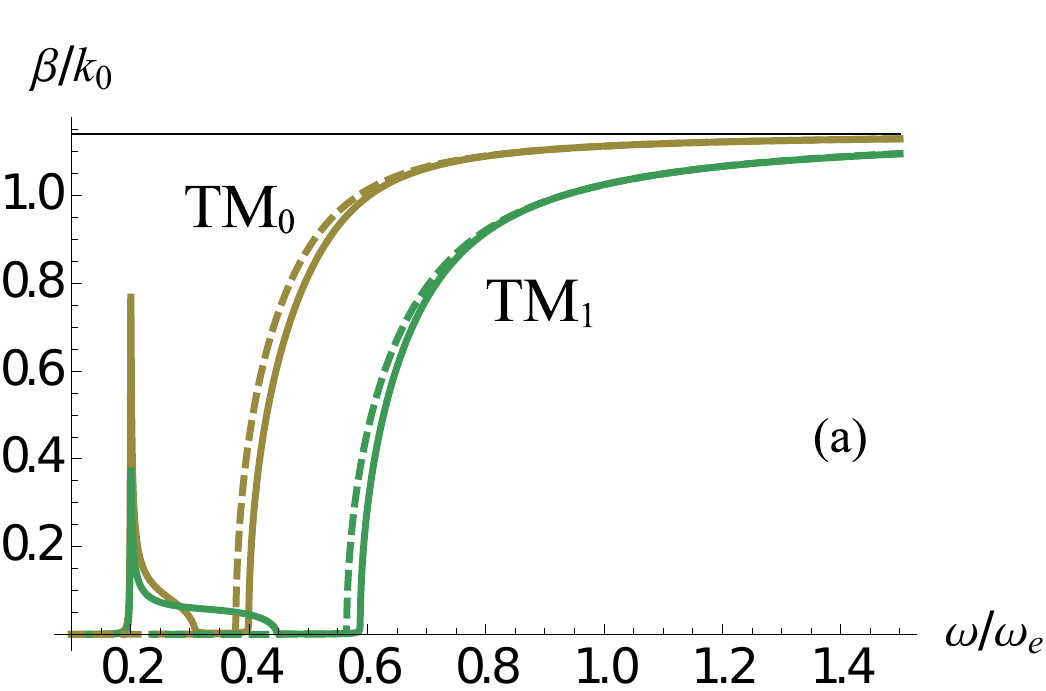}}\\
       \subfloat{\label{fig:TMSlabDispMIXED}\includegraphics[width=0.45\columnwidth]{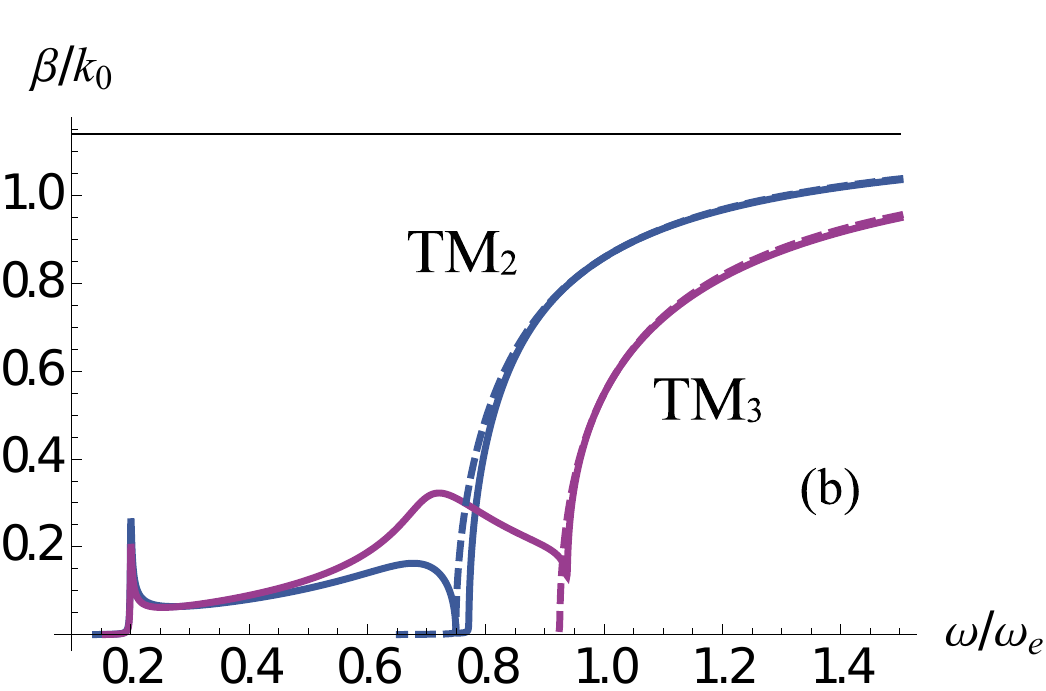}}
	\subfloat{\label{fig:TMSlabDispSurface}\includegraphics[width=0.45\columnwidth]{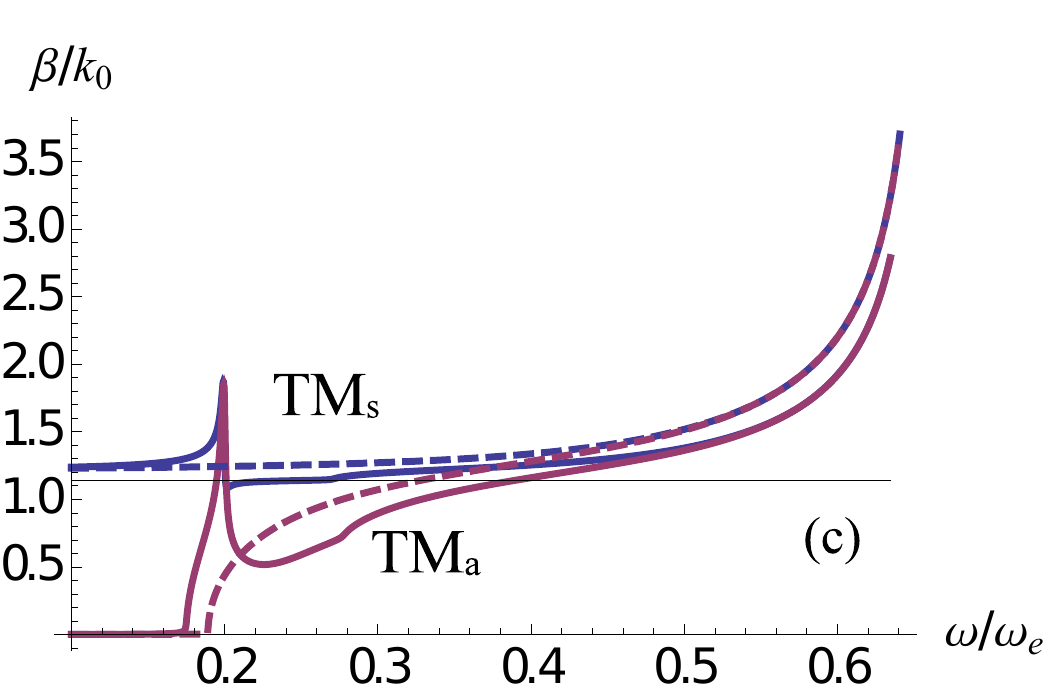}}
\caption{Effective refractive index for \protect\subref{fig:TMSlabDispBOUND} the TM$_{0}$ and TM$_{1}$, \protect\subref{fig:TMSlabDispMIXED} the TM$_{2}$ and TM$_{3}$, and \protect\subref{fig:TMSlabDispSurface} the TM$_{\rm{s}}$ and TM$_{\rm{a}}$ modes modes of the slab guide. The solid lines are for the modes of the metamaterial-dielectric guide, and the dashed lines are for the modes of the metal-dielectric guide. The thin horizontal line in each is the refractive index of the core.}
\label{fig:TMSlabDisp}   
\end{figure}

All guided modes in a waveguide experience an effective refractive index along the propagation direction. The effective refractive index of a waveguide is related to the propagation constant, $\beta$, by the relation $n_{\rm{eff}}=\beta/k_{0}$ with $k_{0}=\omega/c$ the free-space wave number, and c is the speed of light in vacuum. The effective refractive index of ordinary waves in a metamaterial-dielectric guide is very similar to that of a dielectric-dielectric guide, increasing with the frequency of the wave and bound from above by the refractive index of the core, $n_{\rm{core}}=\frac{1}{c}\sqrt{\epsilon_{1}\mu_{1}}$. The effective refractive index of the TM$_{0}$ and TM$_{1}$ modes is plotted in Fig.~\ref{fig:TMSlabDisp}\subref{fig:TMSlabDispBOUND}. For frequencies above $0.3\omega_{\rm e}$ and $0.45\omega_{\rm e}$ the TM$_{0}$ and TM$_{1}$ modes, respectively, are ordinary waves.

Modes with surface-wave behavior exist in metal-dielectric guides as well as in metamaterial-dielectric guides. The requirement of $\epsilon'$ or $\mu'$ to have opposite signs on either side of the interface means surface modes are not possible in dielectric-dielectric guides. Only two modes in the slab guide show surface-wave characteristics, TM$_{\rm{s}}$ and TM$_{\rm{a}}$, which is present at frequencies such that $n_{\rm{eff}}>n_{\rm{core}}$. The effective refractive index of surface waves in a metamaterial-dielectric guide behaves as for a metal-dielectric guide, increasing with the frequency, and is plotted for the slab-guide modes TM$_{\rm{s}}$ and TM$_{\rm{a}}$ in Fig.~\ref{fig:TMSlabDisp}\subref{fig:TMSlabDispSurface}.

Hybrid waves sometimes exhibit anomalous dispersion (decreasing $n_{\rm{eff}}$ with increasing frequency). We have already seen hybrid-mode behavior in Fig.~\ref{fig:TMSlabDisp}\subref{fig:TMSlabDispBOUND}, where it is present at lower frequencies than the ordinary wave of the same mode. There are also regions of hybrid-wave behavior in the TM$_{\rm{s}}$ and TM$_{\rm{a}}$ modes, shown in Fig.~\ref{fig:TMSlabDisp}\subref{fig:TMSlabDispSurface}, which occur for $0.2\omega_{\rm{e}}\lesssim\omega\lesssim0.3\omega_{\rm{e}}$ for the TM$_{\rm s}$ mode and for $0.2\omega_{\rm e}\lesssim\omega\lesssim0.4\omega_{\rm{e}}$ for the TM$_{\rm{a}}$ mode. 

Figure~\ref{fig:TMSlabDisp}\subref{fig:TMSlabDispMIXED} shows the effective refractive index for the TM$_{2}$ and TM$_{3}$ modes of the metamaterial-dielectric slab waveguide. The hybrid-wave behavior occurs for $0.2\omega_{\rm e}\lesssim\omega\lesssim0.75\omega_{\rm e}$ in the TM$_{2}$ mode and $0.2\omega_{\rm e}\lesssim\omega\lesssim0.95\omega_{\rm e}$ in the TM$_{3}$ mode. Both modes have regions of anomalous dispersion with a particularly large frequency region of anomalous dispersion in the TM$_{3}$ mode. 

\begin{figure}[t,b] 
     \centering
       \subfloat{\label{fig:TMSlabHeightBOUND}\includegraphics[width=0.45\columnwidth]{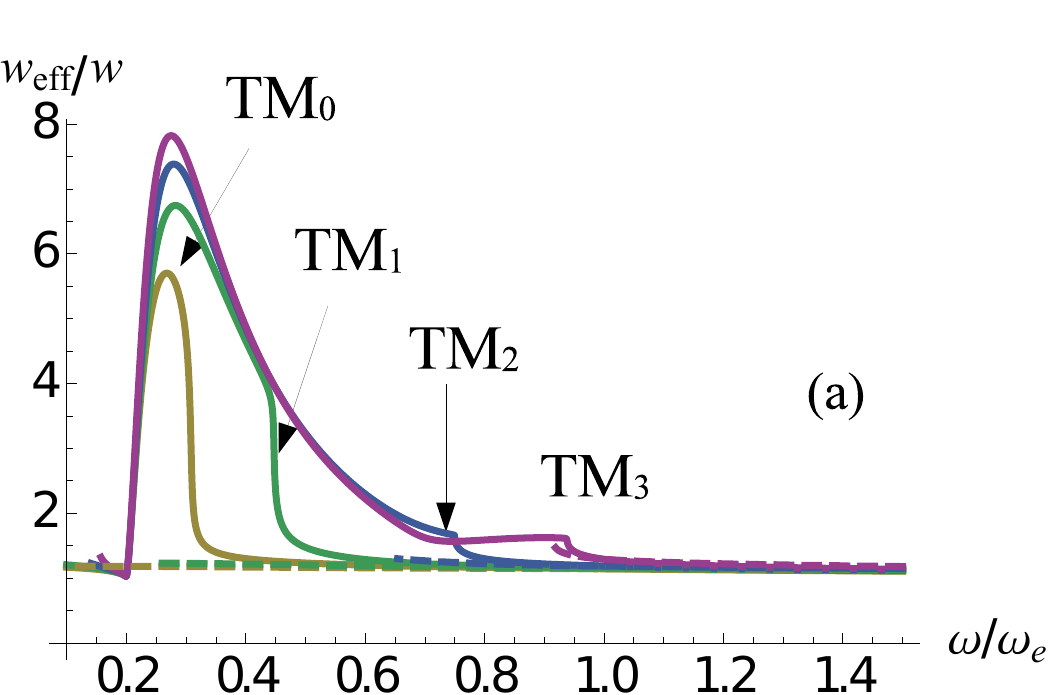}}
       \subfloat{\label{fig:TMSlabHeightSURFACE}\includegraphics[width=0.45\columnwidth]{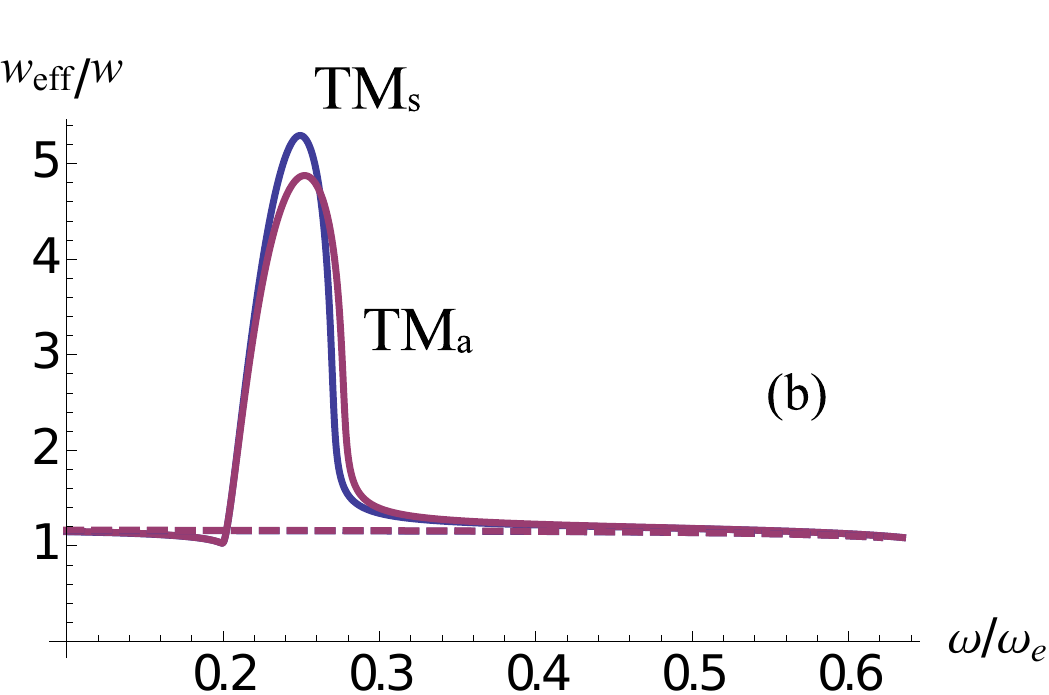}}
     \caption{Effective guide width for \protect\subref {fig:TMSlabHeightBOUND} the TM$_{0}$ to TM$_{3}$, and \protect\subref{fig:TMSlabHeightSURFACE} the TM$_{\rm{s}}$ and TM$_{\rm{a}}$ modes of the slab guide. Solid lines are for the metamaterial-dielectric guide, and dashed lines are for the metal-dielectric guide.}\label{fig:TMSlabHeight}   
\end{figure}

The effective guide width of a waveguide, $w_{\rm{eff}}$, is the nominal width, $w$, plus the skin-depth at each interface, where $w_{\rm{eff}}=w+1/\Re{\gamma_2}+1/\Re{\gamma_3}$ for the slab guide. Ordinary and surface waves in a metamaterial-dielectric guide have an effective width that is only slightly larger than the nominal width; this is consistent with ordinary and surface waves of a metal guide. Hybrid waves, however, display an increased effective width when compared to the ordinary and surface waves. Figures~\ref{fig:TMSlabHeight}\subref{fig:TMSlabHeightBOUND} and \ref{fig:TMSlabHeight}\subref{fig:TMSlabHeightSURFACE} are plots of the relative effective width $w_{\rm{eff}}/w$ and show the large effective width of the hybrid waves.
The largest effective width for each mode is seen in the negative index region, where all modes are hybrid waves. Significantly increased effective width is a feature unique to metamaterial guides. The effective width of metal-dielectric and dielectric-dielectric guides does not have such a large variation.

A large effective width is due to the field having a large extent in the cladding, meaning a significant fraction of the energy is carried outside the core. As the cladding dissipates energy, there is a connection between effective width and losses. The surface waves, however, can have a small effective width even when there is a large fraction of the total energy in the cladding, which is due to the fact that the energy of a surface mode is concentrated around the interface with almost no energy in the center of the core. Thus, even when there is little penetration of the fields into the cladding, there can be a comparably small penetration away from the interface into the core. The effect, then, is a relatively even distribution of energy about the interface and a small effective width. 

The attenuation of a mode in a waveguide is a measure of how much the intensity of the mode decreases as a function of the distance travelled. The imaginary part of $k_{\rm{z}}$ is the attenuation constant, denoted $\alpha$, and characterizes the amount of attenuation per unit length that affects a propagating mode. All of the mode types in the metamaterial guide show some amount of attenuation, with some wave types having more than others. 

\begin{figure}[t,b] 
       \centering
   \subfloat{\label{fig:TMSlabAbsMIXED}\includegraphics[width=0.45\columnwidth]{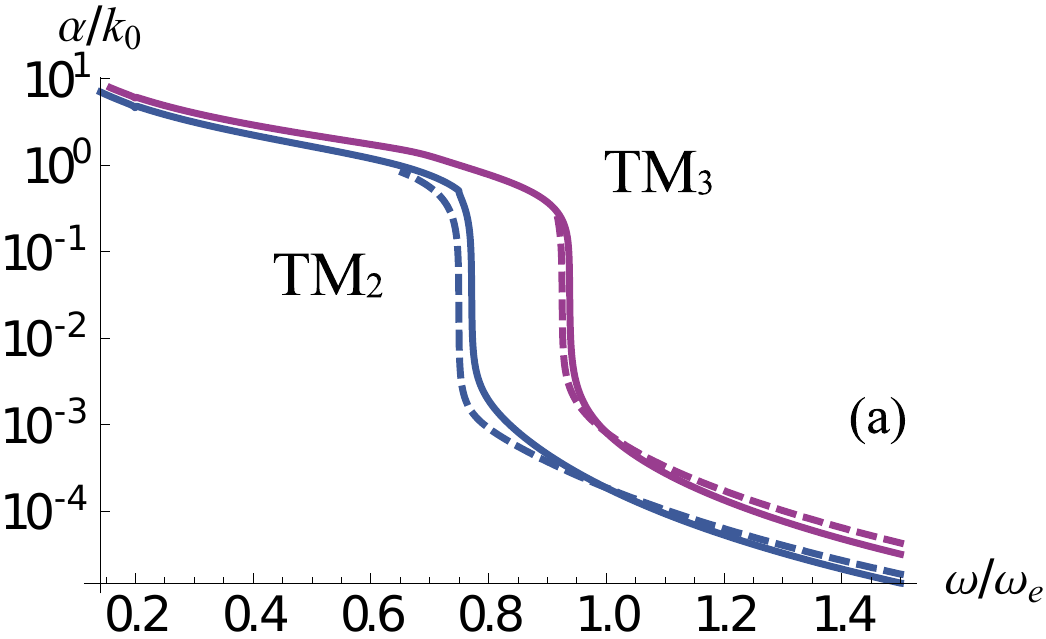}}
   \subfloat{\label{fig:TMSlabAbsSurface}\includegraphics[width=0.45\columnwidth]{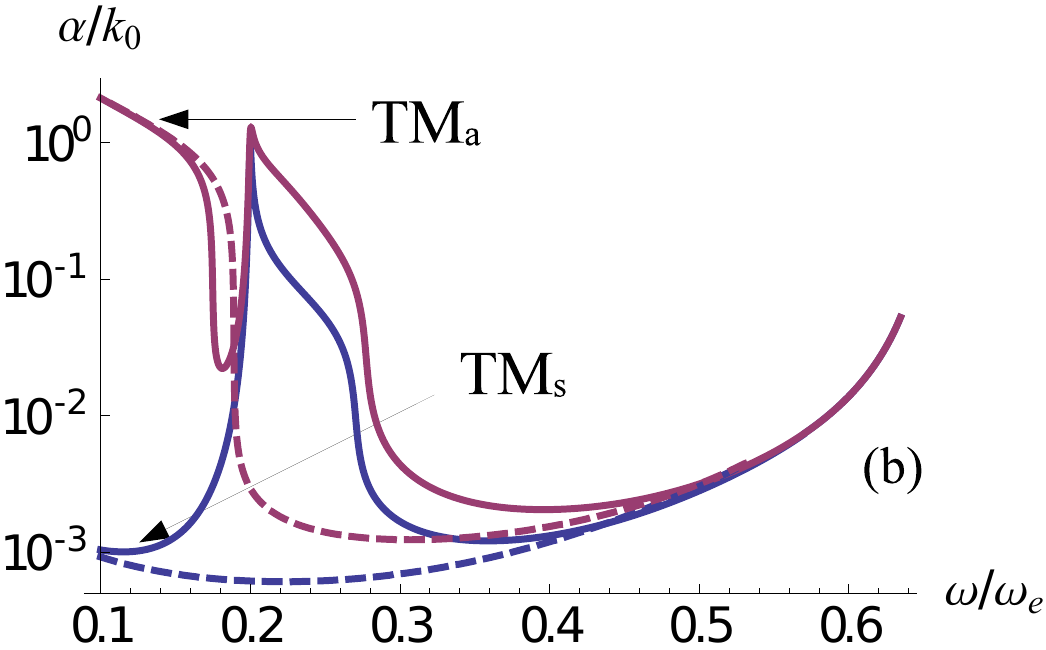}}
     \caption{Attenuation for \protect\subref{fig:TMSlabAbsMIXED} the TM$_{2}$ and TM$_{3}$, and \protect\subref{fig:TMSlabAbsSurface} the TM$_{\rm{s}}$ and TM$_{\rm{a}}$ modes of the slab guide. Solid lines are for the metamaterial-dielectric guide, and dashed lines are for the metal-dielectric guide.}\label{fig:TMSlabAbs}
\end{figure}

Hybrid waves have the largest attenuation, which is partly due to their large effective width. The attenuation of the TM$_{2}$ and TM$_{3}$ modes of the slab guide is be seen in Fig.~\ref{fig:TMSlabAbs}\subref{fig:TMSlabAbsMIXED}.
The regions of large attenuation occur where the modes display hybrid-wave behavior. Hybrid waves also have larger attenuation than surface waves in the same mode, though the difference is much less pronounced. The attenuation of the TM$_{\rm{s}}$ and TM$_{\rm{a}}$ modes is shown in Fig.~\ref{fig:TMSlabAbs}\subref{fig:TMSlabAbsSurface}.

At frequencies where the mode displays hybrid-wave behavior, the attenuation is considerably larger than for either ordinary-wave or surface-wave behavior. The exception is the TM$_{\rm{a}}$ mode for $\omega\lesssim 0.2\omega_{\rm{e}}$ where it is an ordinary wave. Here the attenuation is large also but is consistent with the TM$_{\rm{a}}$ mode in the metal guide (see the dashed line in Fig.~\ref{fig:TMSlabAbs}\subref{fig:TMSlabAbsSurface}), which is also an ordinary wave at low frequencies. The attenuation of both ordinary and hybrid waves decreases as the frequency increases, whereas the attenuation of surface waves generally increases with increasing frequency.

\begin{figure}[t,b] 
     \centering
      \subfloat{\label{fig:relgammaCylOrd1}\includegraphics[width=0.45\columnwidth]{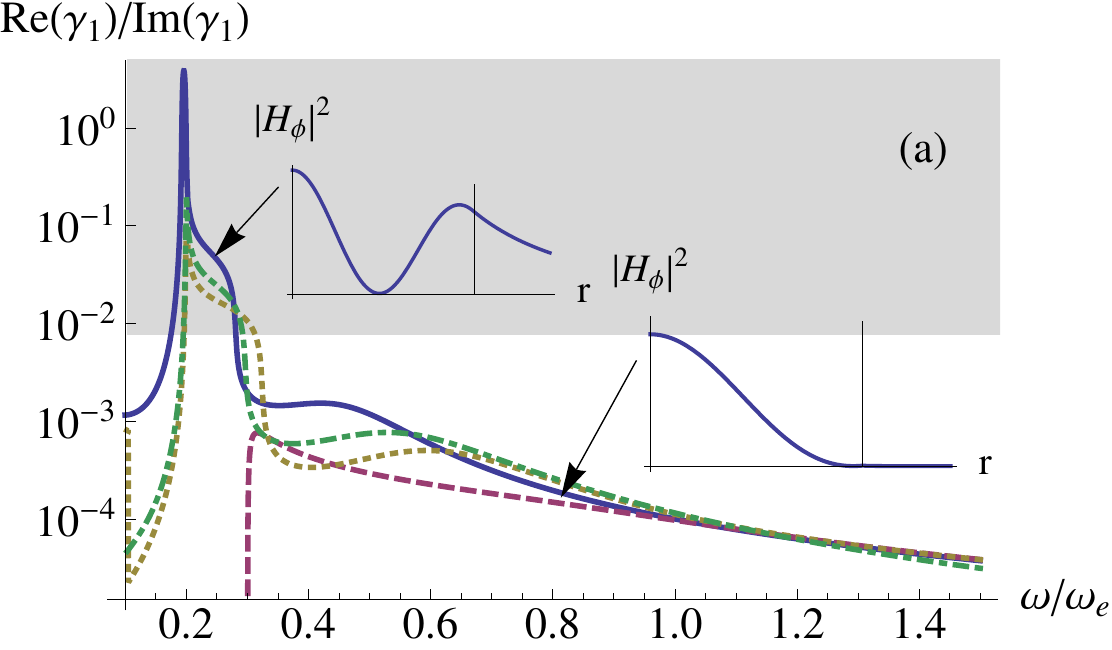}}\\
      \subfloat{\label{fig:relgammaCylOrd2}\includegraphics[width=0.45\columnwidth]{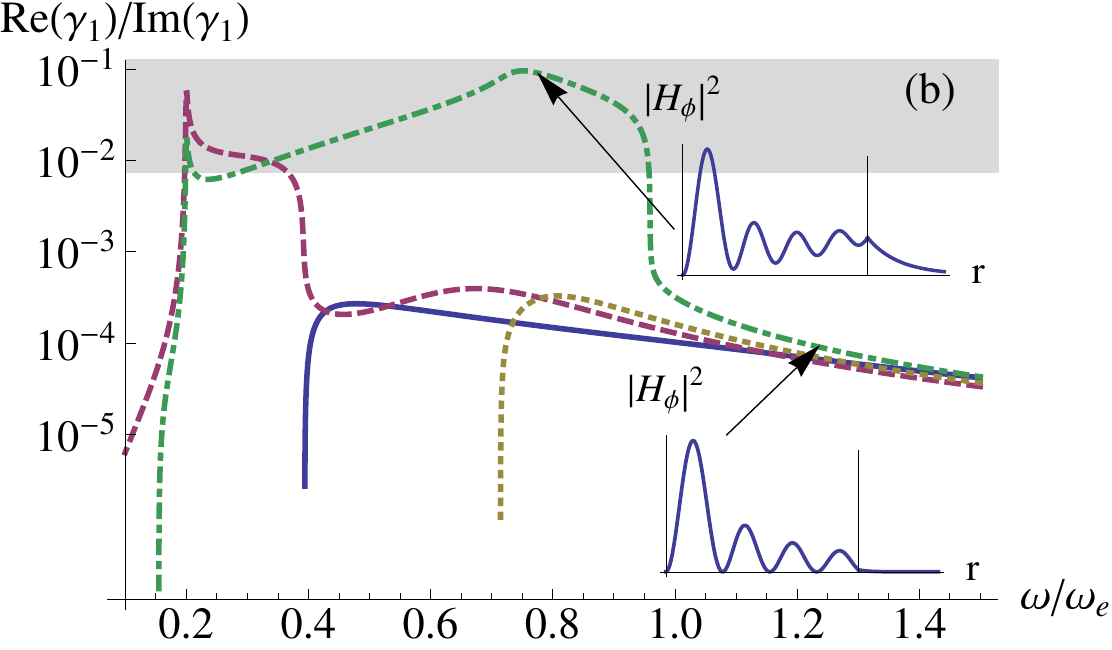}}
      \subfloat{\label{fig:relgammaCylSurf}\includegraphics[width=0.45\columnwidth]{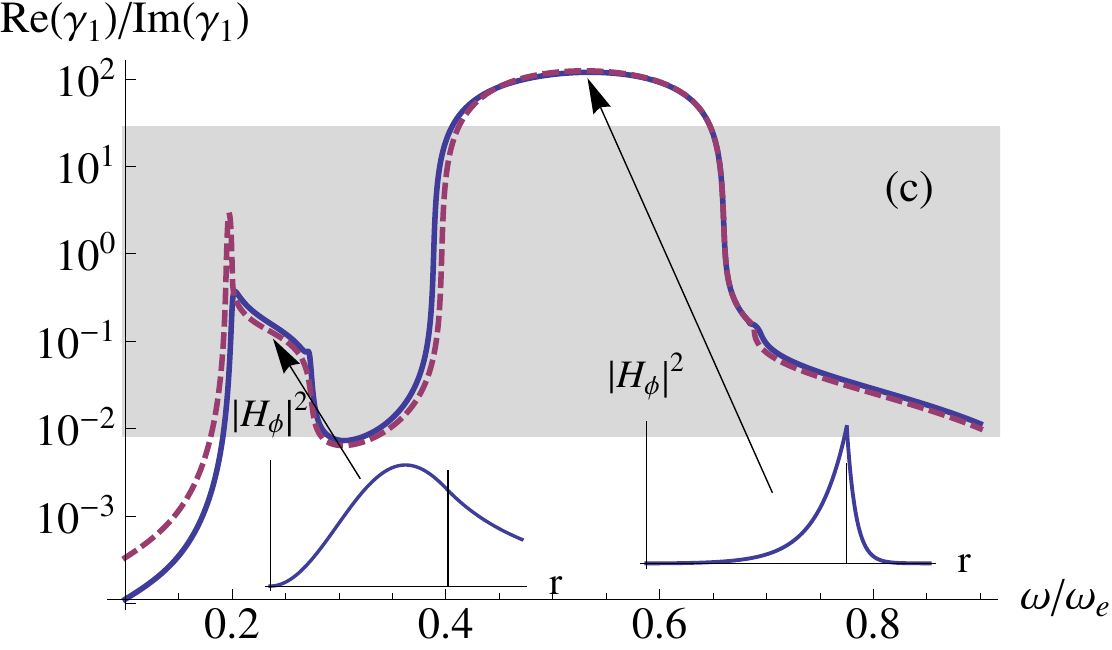}}
     \caption{Plots of $\Re{\gamma_{1}}/\Im{\gamma_{1}}$ for \protect\subref {fig:relgammaCylOrd1} the HE$_{10}$ (solid), HE$_{11}$ (dashed), HE$_{12}$ (dotted), and TM$_{0}$, (dot-dashed), \protect\subref{fig:relgammaCylOrd2} the HE$_{13}$ (solid), TM$_{1}$ (dashed), TM$_{2}$ (dotted), and TM$_{3}$, (dot-dashed) and \protect\subref{fig:relgammaCylSurf} the HE$_{1\rm{s}}$ (solid) and TM$_{\rm{s}}$ (dashed) modes of the cylindrical metamaterial-dielectric guide. The shaded regions indicate the nominal range of hybrid waves, as there is no well-defined boundary. The insets are plots of $|H_{\phi}|^{2}$ for half of the guide at select frequencies and show the transverse profile of the mode at these frequencies. The thin vertical bar in the insets indicates the core/cladding boundary.}\label{fig:relgammaCyl}
\end{figure}
As with the slab guide, we provide plots of $\Re{\gamma_{1}}/\Im{\gamma_{1}}$ for the cylindrical metamaterial-dielectric guide in Fig.~\ref{fig:relgammaCyl} to clarify where the modes are hybrid waves. The fat tails indicative of hybrid waves in a slab guide are also present for hybrid waves in a cylindrical guide. The transverse profiles of the hybrid waves in Figs.~\ref{fig:relgammaSlab} and \ref{fig:relgammaCyl} show this increased field penetration, which is also reflected in the effective guide width for the modes.  The plots in Fig.~\ref{fig:relgammaCyl} show at which frequencies the modes of the cylindrical metamaterial-dielectric guide display which behavior.

Figure~\ref{fig:relgammaCyl}\subref{fig:relgammaCylOrd1} shows plots of $\Re{\gamma_{1}}/\Im{\gamma_{1}}$ for the modes HE$_{10}$ to HE$_{12}$ and TM$_{0}$. From these plots the different frequency regions exhibit where the modes display hybrid and ordinary-wave behavior. What is immediately clear is that there are no frequencies where the HE$_{11}$ mode is a hybrid wave. Instead it has a cut-off frequency near $0.3\omega_{\rm e}$. The three other modes (HE$_{10}$, HE$_{12}$ and TM$_{0}$) are hybrid waves for $0.2\omega_{\rm e}\lesssim\omega\lesssim 0.3\omega_{\rm{e}}$. Beyond about $0.3\omega_{\rm e}$ all four modes are ordinary waves. The insets in Fig.~\ref{fig:relgammaCyl}\subref{fig:relgammaCylOrd1} are $|H_{\phi}|^{2}$ for the HE$_{10}$ mode and show the transverse mode profile at two frequencies, $\omega\approx0.8\omega_{\rm e}$, where the mode is an ordinary wave, and $\omega\approx0.25\omega_{\rm e}$, where the mode is a hybrid wave. 

Similarly, Fig.~\ref{fig:relgammaCyl}\subref{fig:relgammaCylOrd2} shows plots of $\Re{\gamma_{1}}/\Im{\gamma_{1}}$ for the modes HE$_{13}$ and TM$_{1}$ to TM$_{3}$. Two of the modes (HE$_{13}$ and TM$_{2}$) do not show hybrid-wave behavior, whereas the TM$_{1}$ and TM$_{3}$ are hybrid waves for $0.2\omega_{\rm e}\lesssim\omega\lesssim 0.4\omega_{\rm{e}}$ and $0.2\omega_{\rm e}\lesssim\omega\lesssim 0.95\omega_{\rm{e}}$ respectively. The insets here are plots of  $|H_{\phi}|^{2}$ for the TM$_{3}$ mode at two frequencies. Again, one of these plots represents an ordinary mode, $\omega\approx1.2\omega_{\rm e}$, and the other represents a hybrid mode, $\omega\approx0.8\omega_{\rm e}$. 

Modes with surface-wave behavior are supported by the metamaterial-dielectric cylindrical guide as well. Plots of $\Re{\gamma_{1}}/\Im{\gamma_{1}}$ for the HE$_{1\rm{s}}$ and TM$_{\rm s}$ modes, both of which are surface waves for $0.4\omega_{\rm e}\lesssim\omega\lesssim 0.65\omega_{\rm{e}}$, are shown in Fig.~\ref{fig:relgammaCyl}\subref{fig:relgammaCylSurf}. These modes are hybrid waves otherwise, except for $\omega\lesssim0.2\omega_{\rm e}$ where they are ordinary waves. The insets in Fig.~\ref{fig:relgammaCyl}\subref{fig:relgammaCylSurf} show plots of $|H_{\phi}|^{2}$ for the TM$_{s}$ mode showing a surface wave, $\omega\approx0.5\omega_{\rm e}$, and a hybrid wave, $\omega\approx0.25\omega_{\rm e}$. 

\begin{figure}[t,b] 
     \centering
       \subfloat{\label{fig:TMCylDispBOUND}\includegraphics[width=0.45\columnwidth]{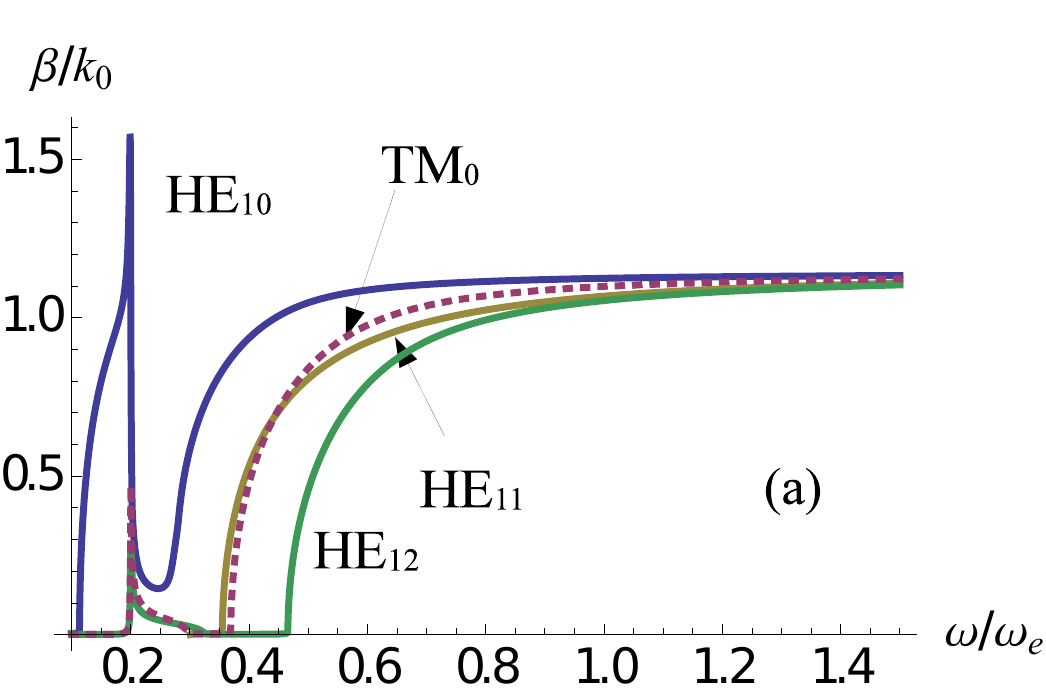}}\\
       \subfloat{\label{fig:CylDispOrdinary5-8}\includegraphics[width=0.45\columnwidth]{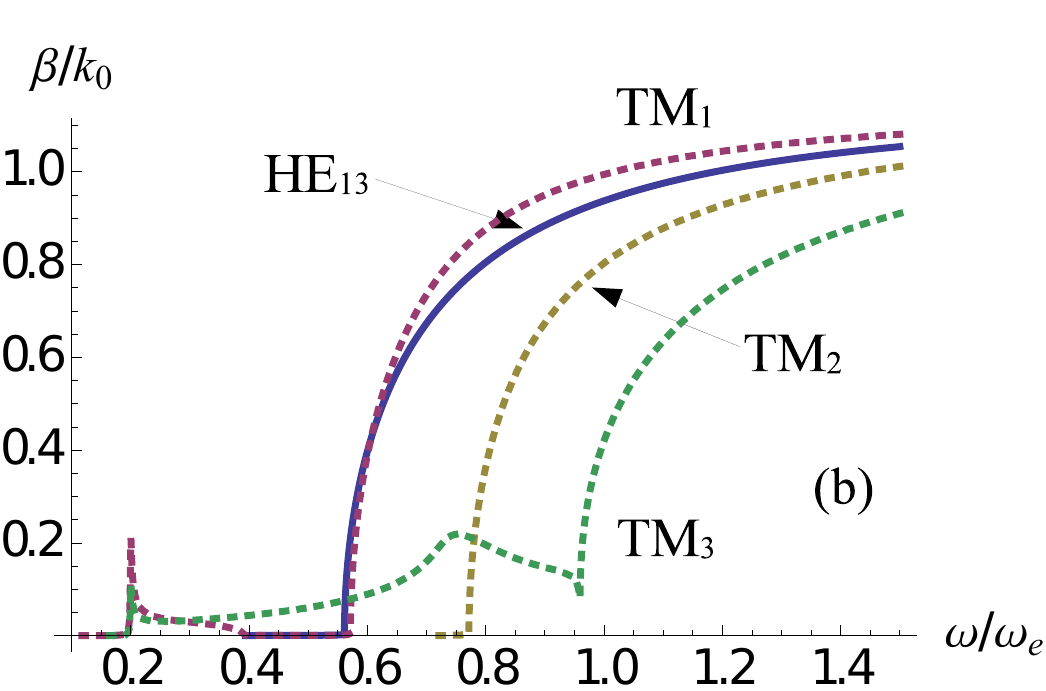}}
       \subfloat{\label{fig:HE1CylDispSURFACE}\includegraphics[width=0.45\columnwidth]{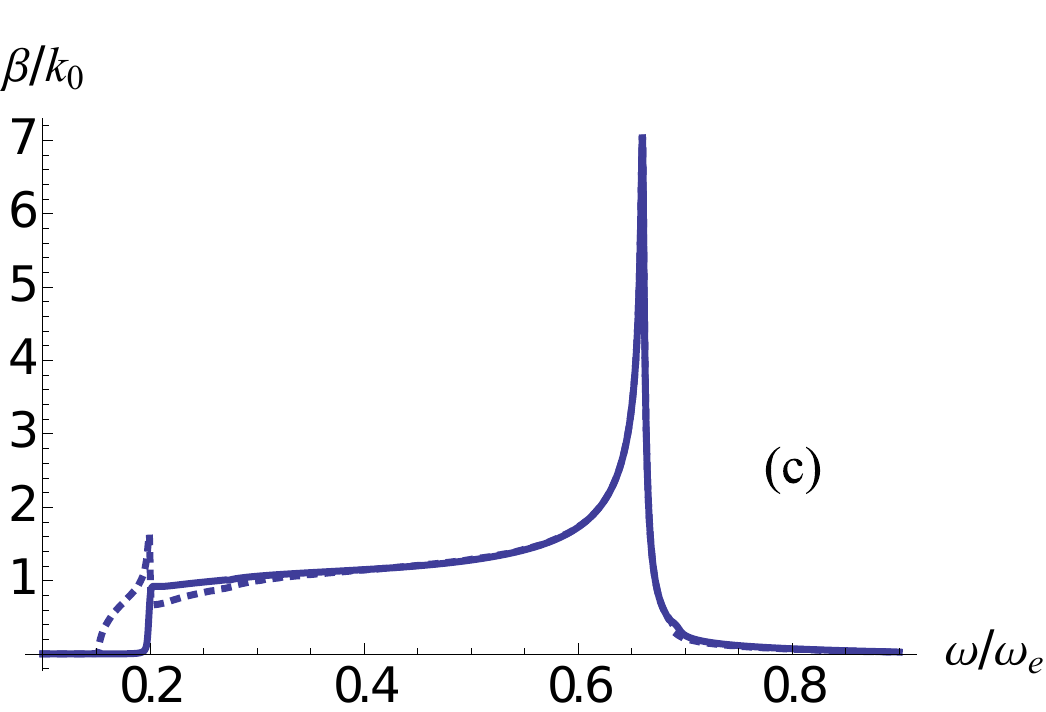}}
     \caption{Effective refractive index for \protect\subref{fig:TMCylDispBOUND} the HE$_{10}$ to HE${12}$, and TM${0}$, \protect\subref{fig:CylDispOrdinary5-8}the  HE$_{13}$, and TM${1}$ to TM${3}$, and \protect\subref{fig:HE1CylDispSURFACE} the HE$_{1\rm{s}}$ and TM$_{\rm{s}}$ modes of the metamaterial-dielectric cylindrical guide. The solid lines indicate HE modes and the dotted lines indicate TM modes.}\label{fig:TMCylDisp}   
\end{figure}

The modes of a cylindrical guide display similar dispersion characteristics to those in a slab guide, including the appearance of hybrid waves at lower frequencies. Figures~\ref{fig:TMCylDisp}\subref{fig:TMCylDispBOUND} to \subref{fig:HE1CylDispSURFACE}
show $n_{\rm{eff}}$ for the modes HE$_{10}$ to HE$_{13}$, TM$_{0}$ to TM$_{3}$, as well as HE$_{\rm s}$, and TM$_{\rm s}$ of the metamaterial cylindrical guide. In contrast to the slab guide, which supports only two surface modes, the cylindrical guide supports a number of modes with surface-wave behavior (one TM and the rest HE), all of different orders (number of oscillations in the electric field over an angle of $2\pi$). 

\begin{figure}[t,b] 
     \centering
     \subfloat{\label{fig:CylHeightOrdinary5-8}\includegraphics[width=0.45\columnwidth]{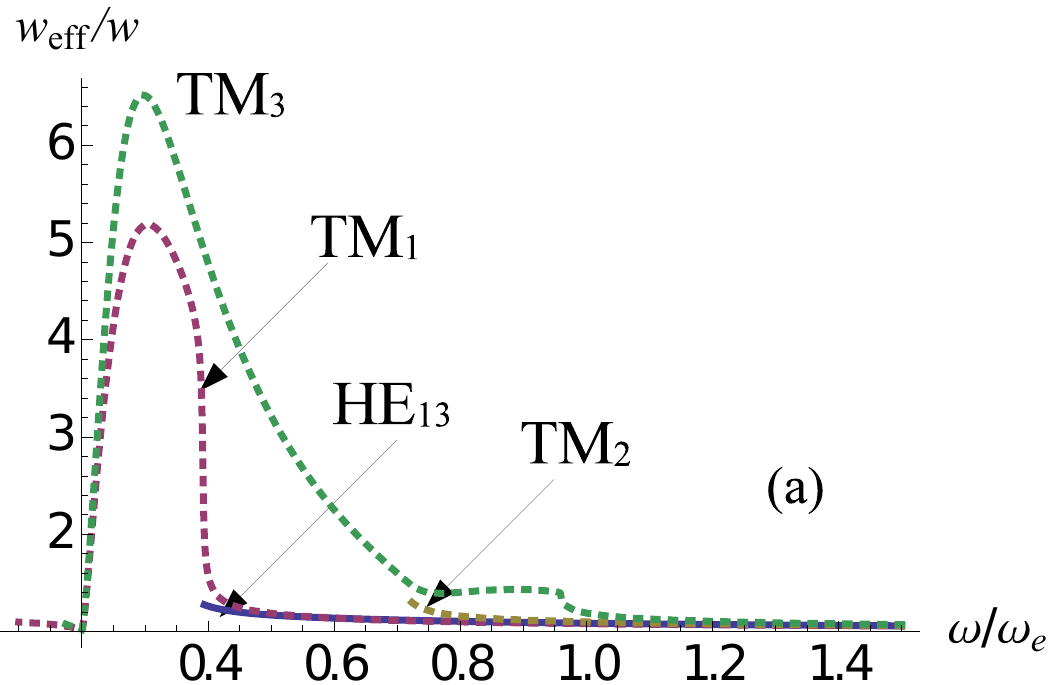}}
     \subfloat{\label{fig:CylHeightSurface}\includegraphics[width=0.45\columnwidth]{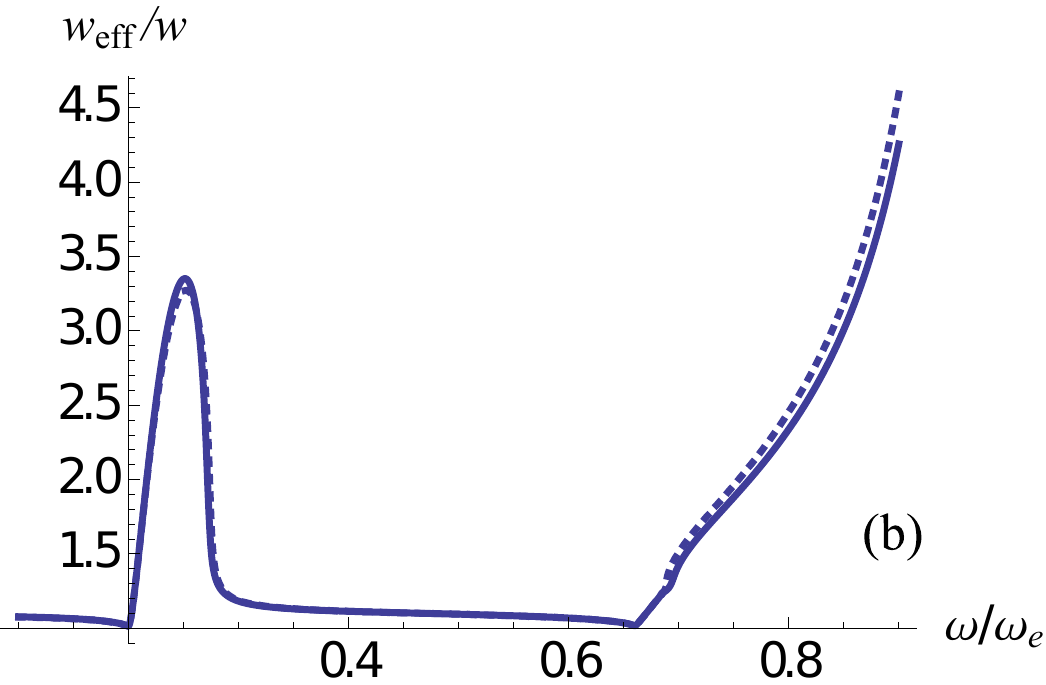}}
  \caption{Effective guide width for \protect\subref{fig:CylHeightOrdinary5-8} the HE$_{13}$ mode, and TM$_{1}$ to TM$_{3}$, and \protect\subref{fig:CylHeightSurface} the HE$_{1\rm{s}}$ and TM$_{\rm{s}}$ modes of the metamaterial-dielectric cylindrical guide. The solid lines indicate HE modes and the dotted lines indicate TM modes.}\label{fig:CylHeight}   
\end{figure}

The effective widths of the modes supported by the metamaterial-dielectric cylindrical guide show similar characteristics to the effective widths of the modes in a metamaterial-dielectric slab guide. Figure~\ref{fig:CylHeight}\subref{fig:CylHeightOrdinary5-8}
shows the effective width of the four modes HE$_{13}$, TM$_{1}$, TM$_{2}$, and TM$_{3}$ of the cylindrical guide, and Fig.~\ref{fig:CylHeight}\subref{fig:CylHeightSurface} shows the width for the HE$_{1\rm{s}}$ and TM$_{\rm{s}}$ modes. The effective widths of the ordinary and surface waves are only slightly larger than the nominal width of the guide, whereas the effective widths of the hybrid waves are large, when compared to the nominal guide width and the effective widths of the other wave types.

\begin{figure}[t,b] 
     \centering
     \subfloat{\label{fig:CylAbsOrdinary5-8}\includegraphics[width=0.45\columnwidth]{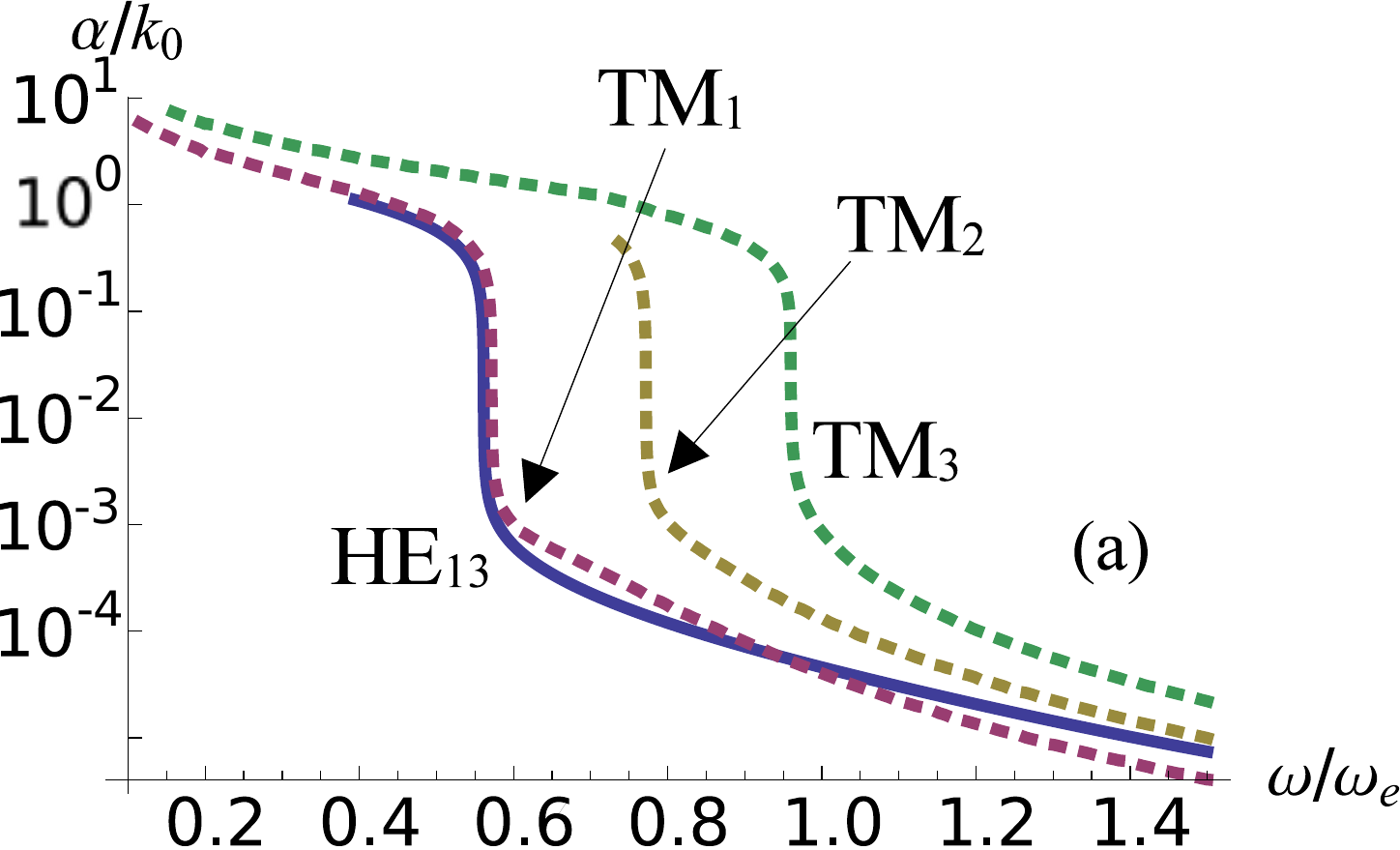}}
     \subfloat{\label{fig:CylAbsSurface}\includegraphics[width=0.45\columnwidth]{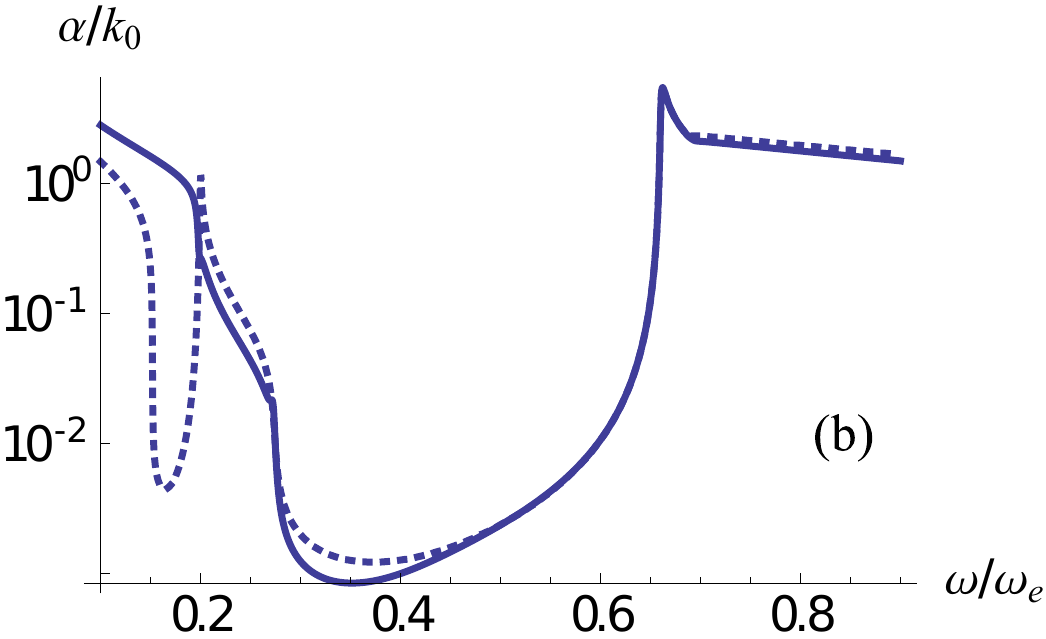}}
  \caption{Attenuation for \protect\subref{fig:CylAbsOrdinary5-8} the HE$_{13}$ and TM$_{1}$ to TM$_{3}$, and \protect\subref{fig:CylAbsSurface} the HE$_{1\rm{s}}$ and TM$_{\rm{s}}$ modes of the metamaterial-dielectric cylindrical guide. The solid lines indicate HE modes and the dotted lines indicate TM modes.}\label{fig:CylAbs}   
\end{figure}

As with the effective widths, the attenuation of the modes of the metamaterial-dielectric cylindrical guide is similar to that of the slab guide. The attenuation for the HE$_{13}$, TM$_{1}$, TM$_{2}$, and TM$_{3}$ modes is shown in Fig.~\ref{fig:CylAbs}\subref{fig:CylAbsOrdinary5-8}.
In frequency regions of hybrid-wave behavior the modes have large attenuation. The same trend of decreasing attenuation with increasing frequency is also present. The TM$_{\rm{s}}$ and HE$_{1\rm{s}}$ modes of the cylindrical guide, shown in Fig.~\ref{fig:CylAbs}\subref{fig:CylAbsSurface}, have attenuation characteristics similar to the TM$_{\rm{a}}$ mode of the slab guide. 

We now understand how metamaterial-dielectric waveguides behave when the effect of absorption is considered. The subject of possible implications and implementations of such waveguides can now be approached.


\section{Discussion: Low-loss surface modes}

In a metamaterial-dielectric guide, attenuation of ordinary and surface waves is much less than for hybrid waves but is still comparable to attenuation in metal-dielectric waveguides. With current metamaterial technology, losses due to atomic interaction with the magnetic field are significant~\cite{Xiao:2009, Penciu:2010}. Previous results for optical metamaterials show magnetic losses could be reduced through structural improvements~\cite{Zhou:2008}. Our results  show that lowering the magnetic losses opens up the possibility of low-loss metamaterial-dielectric waveguides. The attenuation of metamaterial-dielectric waveguides is reduced due to interference effects in the metamaterial~\cite{Kamli:2008,Moiseev:2010}.

\begin{figure}[t,b] 
      \centering
       \subfloat{\label{fig:TMSlabAbsSurfacegm0}\includegraphics[width=0.45\columnwidth]{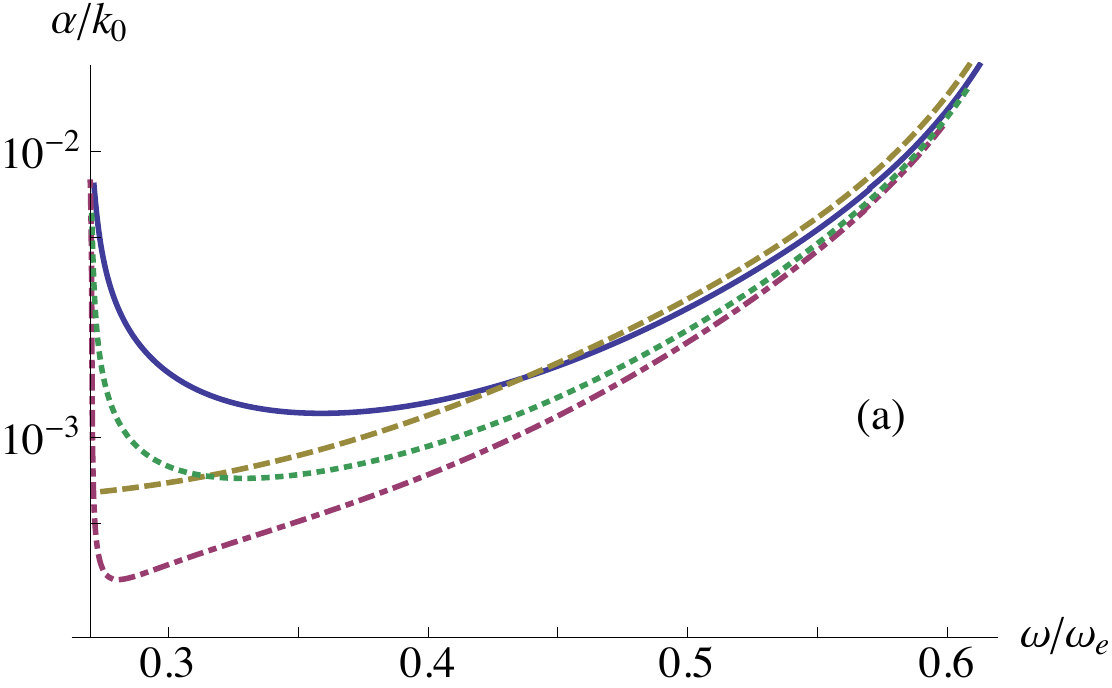}}\\
	\subfloat{\label{fig:HE1CylAbsSurfacegm0}\includegraphics[width=0.45\columnwidth]{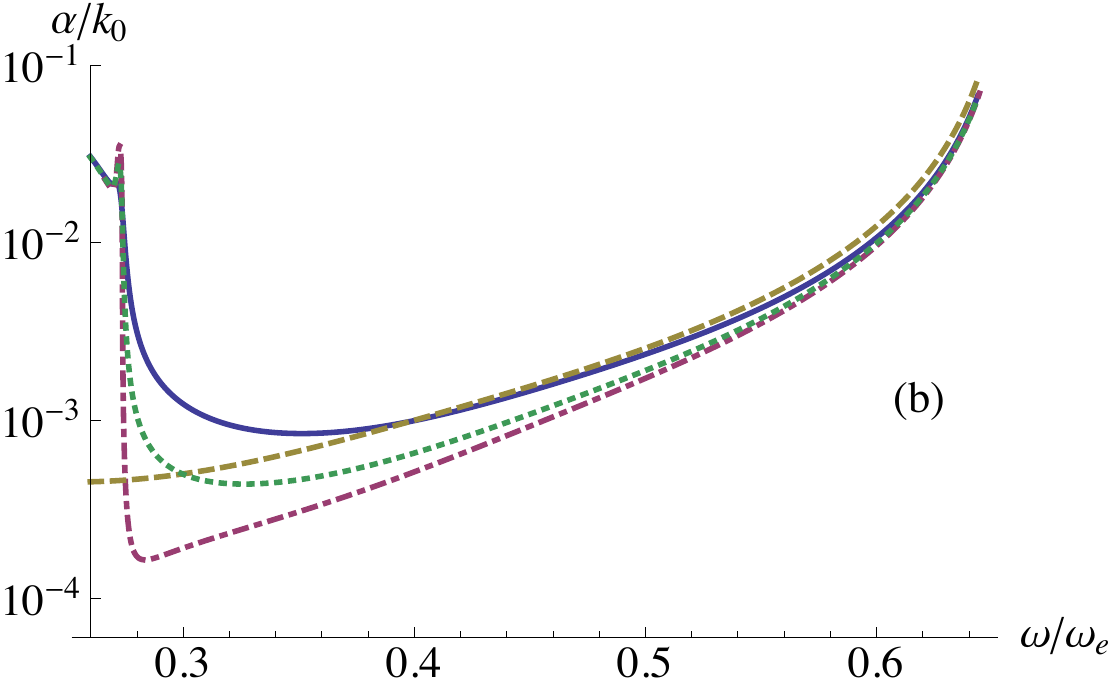}}
	\subfloat{\label{fig:HE1CylHeightSurfacegm0}\includegraphics[width=0.45\columnwidth]{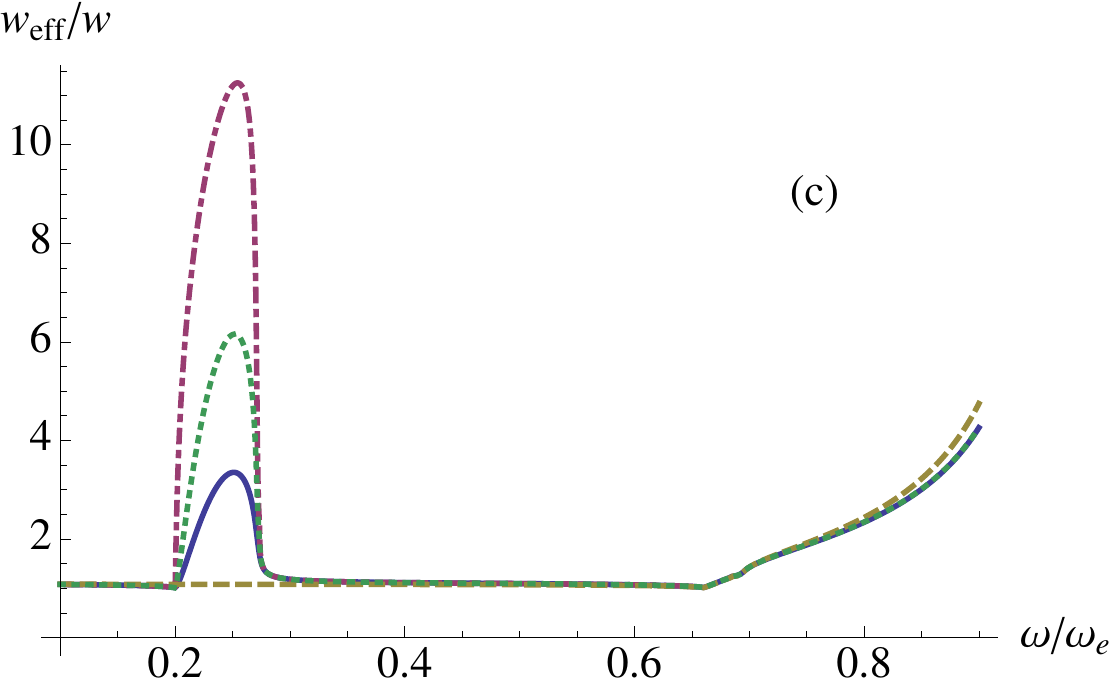}}
	\caption{Attenuation for \protect\subref{fig:TMSlabAbsSurfacegm0} the TM$_{\rm{s}}$ mode of the slab guide and \protect\subref{fig:HE1CylAbsSurfacegm0} the HE$_{\rm{1s}}$ mode of the cylindrical guide, and \protect\subref{fig:HE1CylHeightSurfacegm0} the effective width of the HE$_{\rm{1s}}$ mode of the cylindrical guide. The solid line is for the metamaterial-dielectric guide, the dashed line is for the metal-dielectric guide the dot-dashed line is for a metamaterial-dielectric guide with $\Gamma_{\rm{m}}=\Gamma_{\rm{e}}/100$ for the metamaterial and the dotted line is for a metamaterial-dielectric guide with $\Gamma_{\rm{m}}=\Gamma_{\rm{e}}/3$ for the metamaterial.}
     \label{fig:AbsSurfacegm0}   
\end{figure}

The losses associated with the magnetic interaction are parameterized by $\Gamma_{\rm{m}}$. Figure~\ref{fig:AbsSurfacegm0}\subref{fig:TMSlabAbsSurfacegm0}
shows a comparison of the TM$_{\rm{s}}$ modes for the metamaterial slab guide ($\Gamma_{\rm{m}}=\Gamma_{\rm{e}}=2.73\times10^{13}{\rm s}^{-1}$), the metal slab guide ($\Gamma_{\rm{m}}=0$), the metamaterial slab guide with reduced magnetic losses ($\Gamma_{\rm{m}}=\Gamma_{\rm e}/3$), as well as the metamaterial slab guide with greatly reduced magnetic losses ($\Gamma_{\rm{m}}=\Gamma_{\rm e}/100$). We choose $\Gamma_{\rm{m}}=\Gamma_{\rm{e}}/3$ to generate the plots as this should be practically achievable~\cite{Zhou:2008}.  The attenuation dip seen in metamaterial-dielectric waveguides is related to that seen for surface waves on a single metamaterial-dielectric interface~\cite{Kamli:2008,Moiseev:2010}. The solution for a single interface, as well as the attenuation dip, is recoverable from the metamaterial guides in the limit of large guide width.

Close inspection of Fig.~\ref{fig:AbsSurfacegm0}\subref{fig:TMSlabAbsSurfacegm0} reveals that for $\omega\gtrsim0.45\omega_{\rm e}$ the attenuation of a metamaterial-dielectric guide with $\Gamma_{\rm{m}}=\Gamma_{\rm e}$ is slightly less than that of a metal-dielectric guide. Reducing $\Gamma_{\rm m}$ has the dual effect of further reducing the attenuation of the metamaterial-dielectric guide as well as broadening the frequency window in which the attenuation is less than that of a metal-dielectric guide. Similarly, the attenuation of the HE$_{1\rm{s}}$ mode of the metamaterial-dielectric cylindrical guide has lower attenuation than that for the metal-dielectric guide for $\omega\gtrsim0.45\omega_{\rm e}$. Reducing $\Gamma_{\rm m}$ for the cylindrical guide has the effect of reducing modal attenuation, as it does for the slab guide. Also, when $\Gamma_{\rm{m}}=\Gamma_{\rm e}/100$ the attenuation of the the HE$_{1\rm{s}}$ mode of the metamaterial guide at $\omega\approx0.3\omega_{\rm e}$ is 0.36 times that for the metal guide.

Reducing the value of $\Gamma_{\rm{e}}$, which parameterizes loss due to atomic interactions with the electric field, also has the effect of reducing the total attenuation of the TM and HE modes in a metamaterial-dielectric waveguide. However, reducing $\Gamma_{\rm{e}}$ for a metal-dielectric waveguide also reduces the attenuation of its TM and HE modes. Thus for a metamaterial-dielectric waveguide, reducing $\Gamma_{\rm{e}}$ alone does not yield any benefit over a similar reduction for a metal-dielectric waveguide. The benefit is in combining the reduced $\Gamma_{\rm{e}}$ with a reduced $\Gamma_{\rm{m}}$ to compound the two effects, which significantly reduces attenuation in metamaterial-dielectric waveguides.

All-optical control of low-intensity pulses requires low attenuation and strong transverse field confinement for large cross-phase modulation. The cylindrical metamaterial-dielectric guide must meet the two requirements simultaneously to suffice for all-optical control. The first requirement, low attenuation, allows a low-intensity pulse to pass through the guide without being lost. By extending the results of Kamli, Moiseev and Sanders~\cite{Kamli:2008}, we have shown cylindrical metamaterial-dielectric waveguides can have lower attenuation than metal-dielectric waveguides when the magnetic losses of the metamaterial are reduced. 

The second requirement is strong confinement of the field in the transverse direction (normal to the core/cladding interface). Strong confinement allows the pulse to be confined to a smaller volume, thus increasing the local field intensity and enhancing the nonlinear response. As the low-loss modes of the cylindrical guide are due to field expulsion from the metamaterial cladding, the fields are almost entirely contained in the core. Thus, the metamaterial-dielectric cylindrical waveguide is a good candidate for enhancing cross-phase modulation between low-intensity pulses.

Metamaterial-dielectric waveguides support modes with three distinct regimes, ordinary, surface, and hybrid ordinary-surface waves. The existence of the hybrid wave regime is a direct result of the waveguide dissipating energy. Models for metamaterial waveguides that do not dissipate energy (lossless) do not predict hybrid modes~\cite{Shadrivov:2003,Qing:2005} due to the fact that the wavenumbers in the lossless model are allowed to be either real or imaginary, but not complex. Thus, the supported modes are either surface or ordinary waves, respectively. Modelling the waveguide with a dissipative material allows the wavenumbers to be complex, meaning the modes can have characteristics of both surface and ordinary waves. Hybrid waves in metamaterial-dielectric guides are much more robust than those in a metal-dielectric guide; i.e., they are supported for a much broader frequency range. The wider frequency range may make observing hybrid waves in practice easier for metamaterial guides than for metal guides.

Modes with hybrid-wave behavior also have large attenuation relative to the other supported wave types. One possible use of hybrid modes is frequency filtering to remove unwanted frequencies from pulses. Pulses with frequencies outside of the hybrid wave region are allowed to propagate further, either as an ordinary wave or as a surface wave depending on the mode. Any pulse with frequencies that fall in the region of large attenuation propagate as a hybrid wave and are quickly extinguished.


\section{Summary}

Our results show that reducing the energy loss associated with the magnetic interactions reduces the attenuation of the surface modes in metamaterial-dielectric waveguides. For $\Gamma_{\rm{m}}\lesssim\Gamma_{\rm{e}}$ the attenuation is below that for a metal-dielectric guide, at certain frequencies, despite the metal and metamaterials having identical expressions for the permittivity. Reducing $\Gamma_{\rm m}$ further reduces the attenuation in the metamaterial-dielectric guide as well as broadening the frequency window of reduced attenuation. For example, with our choice of parameters, the attenuation of HE$_{1\rm{s}}$ mode of the metamaterial-dielectric guide can be reduced to as low as about 0.36 times that of the HE$_{1\rm{s}}$ mode in the metal cylindrical guide.

With the use of structural improvements, or some other means of reducing $\Gamma_{\rm{m}}$, it is possible, in principle, to construct metamaterial-dielectric waveguide devices with low losses compared to a metal-dielectric guide. With the capability of transverse confinement while supporting low loss modes, the metamaterial cylindrical guide is a good candidate for enhancing cross-phase modulation and ultimately all-optical control of low-intensity pulses.

We have examined the modal properties of a slab and cylindrical metamaterial-dielectric waveguide using a model that includes energy loss. We have shown that the modes of metamaterial-dielectric waveguides support three distinct regimes, namely ordinary, surface and hybrid, with hybrid modes being a feature unique to models that include energy loss. Ordinary and surface modes are well known and have been studied previously. Hybrid waves, however, have not been previously examined for metamaterial-dielectric waveguides. 

Ordinary waves have the bulk of their energy distributed throughout the core and are the result of internal reflections and interference effects. Surface waves, on the other hand, have their energy concentrated at the interface(s) of the guide, which is due to the fact that the energy is being transported along the guide through the oscillations of surface electrons. Hybrid waves, as their name suggests, show a combination of the two mechanisms and thus their respective energy distributions.

Hybrid waves typically have a large effective width when compared to the other mode types, which implies a large amount of energy is in the cladding. When the energy of a mode is carried in a cladding that dissipates energy, the result is attenuation of the field. Our model allows for a lossy cladding so the increased amount of energy in the cladding for hybrid modes is consistent with the increased attenuation. Within the negative-index frequency range of the metamaterial, all of the modes of the waveguide are hybrid waves and display the associated large attenuation and increased effective guide width.

The sharp change seen in the attenuation curves could be advantageous as a frequency filter. A pulse propagating through a metamaterial guide would have the frequency components in the high-attenuation regions removed through energy dissipation, while the remainder of the pulse continues to propagate. The frequencies at which the attenuation change occurs may be selected through the design of the metamaterial.

\section*{Acknowledgements}

We appreciate valuable discussions with S. A. Moiseev and A. Kamli and financial support from AITF and NSERC. BCS is partially supported by a CIFAR Fellowship.

\bibliographystyle{model1-num-names}

\end{document}